\documentclass[book,numbers]{elsbook-P04}
		
\usepackage{amsmath,amssymb,amsfonts,amsthm,makeidx,graphicx,colortbl}
\usepackage{helvet}

\usepackage{amsmath}    % MARC CHAUMONT mathematical extension
\usepackage{mathtools}  % MARC CHAUMONT for matrix printing
\usepackage{enumitem}   % MARC CHAUMONT for enumerate with label option
\usepackage{hyperref}   % MARC CHAUMONT for better url integration
	
\makeindex

\begin{document}

\Frontmatter

\Mainmatter

%\usepackage{txfonts} -> Supprimé car pause problème

%% Package added in the RunYourChapter.tex (the master document)
%%
%% \usepackage{amsmath}    % MARC CHAUMONT mathematical extension (pmatrix)
%% \usepackage{mathtools}  % MARC CHAUMONT for matrix printing (pmatrix)
%% \usepackage{enumitem}   % MARC CHAUMONT for enumerate with label option
%% \usepackage{hyperref}   % MARC CHAUMONT for better url integration (href)

\begin{frontmatter}%%% Chapter Opener

\chapter[Deep Learning in steganography and steganalysis]{Deep Learning in steganography and steganalysis from 2015 to 2018}\label{chap1}

\author*[1]{Marc CHAUMONT}%

\address[1]{\orgname{Montpellier University, LIRMM (UMR5506) / CNRS, N\^imes University, France}, \orgdiv{LIRMM/ICAR}, \orgaddress{161, rue Ada, 34392 Montpellier, France}}
\address*[3]{Corresponding: \email{marc.chaumont@lirmm.fr}}

\newcolumntype{M}[1]{>{\raggedright}m{#1}}
\vspace{-10cm}\vspace{-1.5cm}
\begin{tabular}{M{8.5cm}}
{\tiny{\bf (final version - October 2019)}. 
%Nook chapter is will app be published as a book by ELSEVIER Inc. This chapter will appear in 2020 in the book titled: ``{\it Digital Media Steganography: Principles, Algorithms, Advances}''. A revised version should be available before the end of 2019. \\
Elsevier Book chapter, Elsevier Inc, Book titled: ``{\it Digital Media Steganography: Principles, Algorithms, Advances}'', Book Editor: M. Hassaballah, 2020.\\
%The final version should be available mid 2020.
}
\end{tabular}
   
%% for running heads
%\authormark{First Author, Second Author and Third Author}
\titlemark{Deep Learning in steganography and steganalysis from 2015 to 2018}
\chaptermark{Deep Learning in steganography and steganalysis}

\MaxmMiniTocnum{A.1}{3.3.3}{}

\minitoc

\makechaptertitle

\begin{abstract}[Abstract]
For almost 10 years, the detection of a hidden message in an image has been mainly carried out by the computation of Rich Models (RM), followed by classification using an Ensemble Classifier (EC). In 2015, the first study using a convolutional neural network (CNN) obtained the first results of steganalysis by Deep Learning approaching the performances of the two-step approach (EC + RM). Between 2015-2018, numerous publications have shown that it is possible to obtain improved performances, notably in spatial steganalysis, JPEG steganalysis, Selection-Channel-Aware steganalysis, and in quantitative steganalysis.
This chapter deals with deep learning in steganalysis from the point of view of current methods, by presenting different neural networks from the period 2015-2018, that have been evaluated with a methodology specific to the discipline of steganalysis. The chapter is not intended to repeat the basic concepts of machine learning or deep learning. So, we will present the structure of a deep neural network, in a generic way and present the networks proposed in existing literature for the different scenarios of steganalysis, and finally, we will discuss steganography by deep learning.
\end{abstract}

\begin{keywords}[Keywords:]
Steganography \sep steganalysis \sep Deep Learning \sep GAN
\end{keywords}

\end{frontmatter}%

Neural networks have been studied since the 1950s. Initially, they were proposed to model the behavior of the brain. In computer science, especially in artificial intelligence, they have been used for around 30 years for learning purposes. Ten or so years ago \cite{HintonSalakhutdinov2006b}, neural networks were considered to have a lengthy learning time and to be less effective than classifiers such as SVMs or random forests.

With recent advances in the field of neuron networks \cite{BengioCV13}, thanks to the computing power provided by graphics cards (GPUs), and because of the profusion of available data, deep learning approaches have been proposed as a natural extension of neural networks. Since 2012, these deep networks have profoundly marked the fields of signal processing and artificial intelligence, because their performances make it possible to surpass current methods, but also to solve problems that scientists had not managed to solve until now\cite{Lecun2015}.

In steganalysis, for the last 10 years, the detection of a hidden message in an image was mainly carried out by calculating Rich Models (RM) \cite{Fridrich2012_Rich} followed by a classification using a classifier (EC) \cite{Kodovsky2012-EnsembleClassifiers}. In 2015, the first study using a convolutional neural network (CNN) obtained the first results of deep-learning steganalysis approaching the performances of the two-step approach (EC + RM~\footnote{We will note EC + RM in order to indicate the two-step approach based on the calculation of Rich Models (RM) then the use of an ensemble classifier (EC).}) \cite{Qian_2015_Deep}. During the period 2015 - 2018, many publications have shown that it is possible to an obtain improved performance in spatial steganalysis, JPEG steganalysis, side-informed steganalysis, quantitative steganalysis, etc.

In Section \ref{sec:briquesCNN} we present the structure of deep neural network generically. This Section is centered on existing publications in steganalysis and should be supplemented by reading about artificial learning and in particular gradient descent, and stochastic gradient descent. In Section \ref{sec:steps_convo} we explain the different steps of the convolution module. In Section \ref{sec:complexity} we will tackle the complexity and learning times. In Section \ref{sec:linktopast} we will present the links between Deep Learning and previous approaches. In Section \ref{sec:periode15-18} we will revisit the different networks that were proposed during the period 2015-2018 for different scenarios of steganalysis. Finally, in Section \ref{sec:GAN} we will discuss steganography by deep learning which sets up a game between two networks in the manner of the precursor algorithm ASO \cite{Kouider2013}.

%%%%%%%%%%%%%%%%%%%%%%%%%%%%%%%%%%%%%%%%%%%%%%%%%%%%%%%%%%%%%%%%%%%%%%%%%%%%%%%%%%%%%%%%%%%%%%%%%%%%%
% SECTION
%%%%%%%%%%%%%%%%%%%%%%%%%%%%%%%%%%%%%%%%%%%%%%%%%%%%%%%%%%%%%%%%%%%%%%%%%%%%%%%%%%%%%%%%%%%%%%%%%%%%%
\section{The building blocks of a deep neuronal network}
\label{sec:briquesCNN}

\begin{figure*}[tb]
  \FIG{\includegraphics[width=\columnwidth]{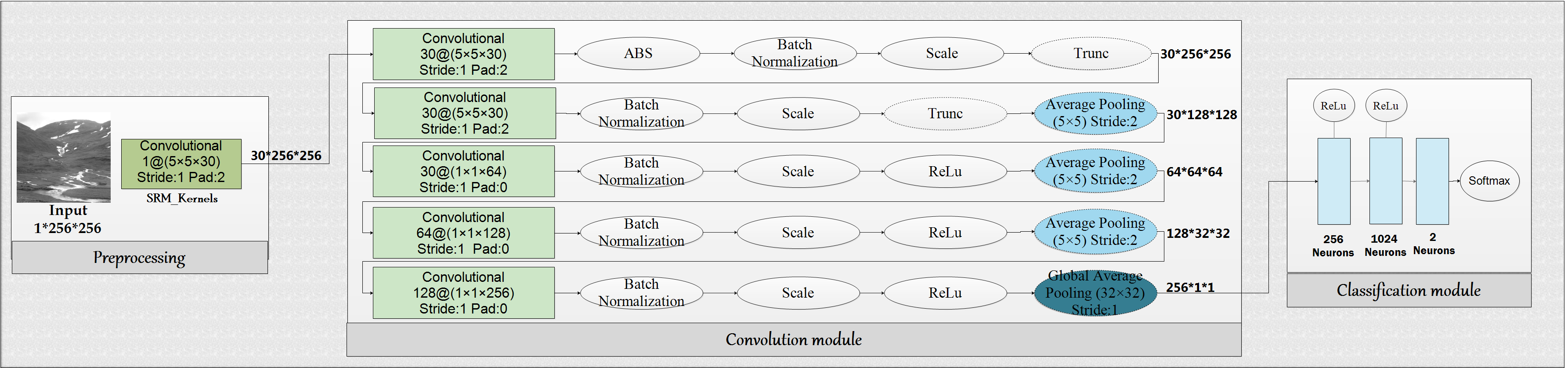}}
		  {\caption{Yedroudj-Net network \cite{Yedroudj2018_Net}.}}
  \label{fig:yed-net}
\end{figure*}

In the following sub-sections, we look back at the major concepts of a Convolutional Neural Network (CNN). But specifically, we will recall the basic building blocks of a network based on the Yedroudj-Net\footnote{GitHub link on Yedroudj-Net: \url{https://github.com/yedmed/steganalysis_with_CNN_Yedroudj-Net}.} network that was published in 2018 \cite{Yedroudj2018_Net} (See Figure \ref{fig:yed-net}), and which takes up the ideas present in Alex-Net \cite{Krizhevsky2012_Convnet}, as well as the concepts present in networks developed for steganalysis including the very first network of Qian {\it et al.} \cite{Qian_2015_Deep}, and networks of Xu-Net \cite{Xu2016a}, and Ye-Net \cite{Ye2017}.

%%%%%%%%%%%%%%%%%%%%%%%%%%%%%%%%%%%%%%%%%%%%%%%%%%%%%%%%%%%%%%%%%%%%%%%%%%%%%%%%%%%%%%%%%%%%%%%%%%%%%
\subsection{Global view of a Convolutional Neural Network}

Before describing the structure of a neural network as well as its elementary components, it is useful to remember that a neural network belongs to the machine-learning family. In the case of supervised learning, which is the case that most concerns us, it is necessary to have a database of images, with, for each image, its label, that is to say, its class.

Deep Learning networks are large neural networks that can directly take raw input data. In image processing, the network is directly powered by the pixels forming the image. Therefore, a deep learning network learns in a joint way, both the compact intrinsic characteristics of an image (we speak of {\it feature map} or of {\it latent space}) and at the same time, the separation boundary allowing the classification (we also talk of {\it separator plans}).

The learning protocol is similar to classical machine learning methods. Each image is given as input to the network. Each pixel value is transmitted to one or more neurons. The network consists of a given number of {\it blocks}. A block consists of neurons that take real input values, perform calculations, and then transmit the actual calculated values to the next block. A neural network can, therefore, be represented by an oriented graph where each node represents a computing unit. The learning is then completed by supplying the network with examples composed of an image and its label, and the network modifies the parameters of these calculation units (it learns) thanks to the mechanism of back-propagation.

The Convolutional Neuronal Networks used for steganalysis are mainly built in three parts, which we will call {\it modules}: the pre-processing module, the convolution module, and the classification module. As an illustration, figure \ref{fig:yed-net} schematizes the network proposed by Yedroudj {\it et al.} in 2018 \cite{Yedroudj2018_Net}. The network processes grayscale images of $ 256\times 256$ pixels.

%%%%%%%%%%%%%%%%%%%%%%%%%%%%%%%%%%%%%%%%%%%%%%%%%%%%%%%%%%%%%%%%%%%%%%%%%%%%%%%%%%%%%%%%%%%%%%%%%%%%%
\subsection{The pre-processing module}

\begin{equation}
F^{(0)} = \frac{1}{12}
\begin{pmatrix*}
   -1 & 2 & -2 & 2 & -1 \\
    2 & -6 & 8 & -6 & 2 \\
   -2 & 8 & -12 & 8 & -2 \\
   2 & -6 & 8 & -6 & 2 \\
   -1 & 2 & -2 & 2 & -1
\end{pmatrix*}
\label{eq:filterF0}
\end{equation}

We can observe in Figure \ref{fig:yed-net} that in the {\it pre-processing module}, the image is filtered by 30 high-pass filters. The use of one or more high-pass filters as pre-processing is present in the majority of networks used for steganalysis during the period 2015-2018. 

\begin{figure*}[ht]
\centering
  \FIG{\includegraphics[width=7cm,keepaspectratio=True]{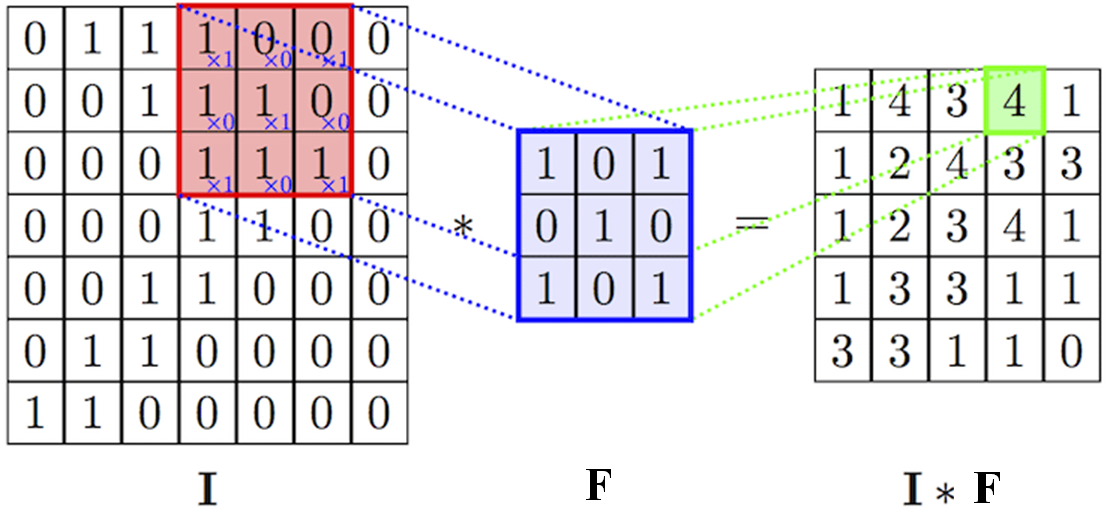}}
      {\caption{Principle of a convolution.}}
  \label{fig:convolution}
\end{figure*}

An example of a kernel of a high-pass filter -- the square  S5a filter \cite{Fridrich2012_Rich} -- is given in Equation \ref{eq:filterF0}. An illustration of the filtering (convolution) principle is given in Figure \ref{fig:convolution}. This preliminary filtering step allows the network to converge faster and is probably needed to obtain good performance when the learning database is too small \cite{Yedroudj2018_DatabaseAugmentation} (only 4 000 pairs cover$/$stego images of size $256\times256$ pixels). The filtered images are then transmitted to the first convolution block of the network. Note that the recent SRNet \cite{Boroumand2018_SRNet} network does not use any fixed pre-filters, but learns the filters. It therefore requires a much larger database (more than 15 000 pairs cover$/$stego images of size $256\times256$ pixels), and strong know-how for its initialization. Note that there is a debate in the community if one should use fixed filters, or initialize the filters with pre-chosen values and then continue the learning, or learn filters with random initialization. At the beginning of 2019, in practice (real-world situation \cite{Ker2013_RealWorld}), the best choice is probably in relation to the size of the learning database (which is not necessary BOSS \cite{Bas2011-BOSS} or BOWS2 \cite{BOWS2008}), and the possibility to use transfer learning.

%%%%%%%%%%%%%%%%%%%%%%%%%%%%%%%%%%%%%%%%%%%%%%%%%%%%%%%%%%%%%%%%%%%%%%%%%%%%%%%%%%%%%%%%%%%%%%%%%%%%%
\subsection{The convolution module}
\label{ssec:module_convolution}

Within the {\it convolution module}, we find several macroscopic computation units that we will call {\it blocks}. A {\it block} is composed of calculation units that take real input values, perform calculations, and return real values, which are supplied to the next block. Specifically, a {\it block} takes a set of {\it feature maps} (= a set of images) as input and returns a set of {\it feature maps} as output (= a set of images). Inside a block, there are a number of operations including the following four: the {\it convolution} (see Section \ref{ssec:convolution}), the {\it activation} (see Section \ref{ssec:activation}), the {\it pooling} (see Section \ref{ssec:pooling}), and finally the {\it normalization} (see Section \ref{ssec:normalization}). 

Note that the concept of neuron, as defined in existing literature, before the emergence of convolutional networks, is still present, but it no longer exists as a data structure in neural network libraries. In convolution modules, we must imagine a neuron as a computing unit which, for a position in the {\it feature map} taken by the convolution kernel during the convolution operation, performs the weighted sum between the kernel and the group of considered pixels. The concept of neuron corresponds to the scalar product between the input data (the pixels) and data specific to the neuron (the weight of the convolution kernel), followed by the application of a function of $\mathbb{R}$ in $\mathbb{R}$ called the activation function. Then, by extension, we can consider that pooling and normalization are operations specific to neurons. 

Thus, the notion of {\it block} corresponds conceptually to a `` layer '' of neurons. Note that in deep learning libraries, we call a {\it layer} any elementary operation such as convolution, activation, pooling, normalization, etc. To remove any ambiguity, for the convolution module we will talk about {\it block}, and {\it operations}, and we will avoid using the term {\it layer}.

Without counting the pre-processing block, the {\it Yedroudj-Net} network \cite{Yedroudj2018_Net} has a convolution module made of 5 convolution blocks, like the networks of Qian {\it et al.} \cite{Qian_2015_Deep} and Xu {\it et al.} \cite {Xu2016a}. The {\it Ye-Net} network \cite{Ye2017} has a convolution module composed of 8 convolution blocks, and SRNet network \cite{Boroumand2018_SRNet} has a convolution module built with 11 convolution blocks. 

%%%%%%%%%%%%%%%%%%%%%%%%%%%%%%%%%%%%%%%%%%%%%%%%%%%%%%%%%%%%%%%%%%%%%%%%%%%%%%%%%%%%%%%%%%%%%%%%%%%%%
\subsection{The classification module}
\label{ssec:classification}

The last block of the convolution module (see the previous Section) is connected to the {\it classification module} which is usually a {\it fully connected} neural network composed of one to three blocks. This {\it classification module} is often a traditional neural network where each neuron is fully connected to the previous {\it block} of neurons and to the next {\it block} of neurons. 

The fully connected blocks often end with a softmax function which normalize the outputs delivered by the network between $[0,1]$, such that the sum of the outputs equal one. The outputs are named imprecisely ``probabilities''. We will keep this denomination. So, in the usual binary steganalysis scenario, the network delivers two values as output: one giving the probability of classifying into the first class (e.g. the cover class), and the other giving the probability of classifying into the second class (e.g. the stego class). The classification decision is then obtained by returning the class with the highest probability.

%-> FAUX !!! Note that in machine-learning \cite{Shelhamer2017_FCN}, and also in steganalysis \cite{Ye2017}, there are networks whose classification module is reduced to a softmax. There is no fully connected block. We then speak of {\it fully convolutional network} \cite{Shelhamer2017_FCN}. One of the advantages of this type of network is that it reduces the number of parameters that the network must learn, which reduces the cost of learning, and thus allows to work with deeper or wider networks. 

Note that in front of this {\it classification module}, we can find a {\it particular pooling} operation such as a {\it global average pooling}, a {\it Spatial Pyramid Pooling (SPP)} \cite {He2014_SPP}, a {\it statistical moments extractor} \cite{Tsang2018_ArbitrarySize}, etc. Such pooling operations return a fixed-size vector of values, that is to say, a feature map of fixed dimensions. The next block to this {\it pooling} operation is thus always connected to a vector of fixed size. So, this block has a fixed input number of parameters. It is thus possible to present to the network images of any size without having to modify the topology of the network. For example, this property is available in the Yedroudj-Net \cite{Yedroudj2018_Net} network, the Zhu-Net \cite{Zhu-Net2019} network, or the {Tsang et al.} network \cite{Tsang2018_ArbitrarySize}.

Also note that \cite{Tsang2018_ArbitrarySize} is the only paper, at the time of writing this chapter, which has seriously considered the viability of an invariant network to the dimension of the input images. The problem remains open. The solution proposed in \cite{Tsang2018_ArbitrarySize} is a variant of the concept of average pooling. For the moment, there has not been enough studies on the subject to determine what is the correct topology of the network, how to build the learning data-base, how much the number of embedded bits influences the learning, or if we should take into account the {\it square root law} for learning at a fixed security-level or any payload size, etc.

%%%%%%%%%%%%%%%%%%%%%%%%%%%%%%%%%%%%%%%%%%%%%%%%%%%%%%%%%%%%%%%%%%%%%%%%%%%%%%%%%%%%%%%%%%%%%%%%%%%%%
% SECTION
%%%%%%%%%%%%%%%%%%%%%%%%%%%%%%%%%%%%%%%%%%%%%%%%%%%%%%%%%%%%%%%%%%%%%%%%%%%%%%%%%%%%%%%%%%%%%%%%%%%%%
\section{The different steps of the convolution module}
\label{sec:steps_convo}

In Section \ref{ssec:module_convolution}, we indicated that a block within the convolution module contained a variable number among the following four operations: the {\it convolution} (see Section \ref{ssec:convolution}), the {\it activation} (see Section \ref{ssec:activation}), the {\it pooling} (see Section\ref{ssec:pooling}), and finally the {\it normalization} (see Section \ref{ssec:normalization}). Let's now explain in more detail each step (convolution, activation, pooling, and normalization) within a {\it block}.

\subsection{Convolution}
\label{ssec:convolution}

The first treatment within a {\it block} is often to apply the convolutions on the input {\it feature maps}.

Note that for the pre-processing {\it block}, see Figure \ref{fig:yed-net}, there is only one input image. A convolution is therefore carried out between the input image and a filter. In the Yedroudj-Net network, there are 30 high-pass filters extracted from SRM filters \cite{Fridrich2012_Rich}. In older networks, there is only one pre-processing filter \cite{Qian_2015_Deep, Pibre2016, Xu2016a}.

Except for the pre-processing {\it block}, in the other {\it blocks}, once the convolution has been applied, we apply activation steps (see Section \ref{ssec:activation}), pooling (see Section \ref{ssec:pooling}), and normalization (see Section \ref{ssec:normalization}). Then we obtain a new image named {\it feature map}.

Formally, let $I^{(0)} $ be the input image of the pre-processing {\it block}. Let $F^{(l)}_k $ be the $k^{th}$ ($k\in\{1, ..., K^{(l)}\}$) filter of the {\it block} of number $l = \{1, ..., L\}$, with $L$ the number of {\it blocks}, and with $K^{(l)}$ the number of filters of the $l^{th}$ {\it block}. The convolution within the pre-processing {\it block} with the $k^{th}$ filter results in a filtered image, denoted $\tilde{I}^{(1)}_k$, such that:

\begin{equation}	
\tilde{I}^{(1)}_k = I^{(0)} \star F^{(1)}_k.
\label{eq:I1}
\end{equation}  

From the first {\it block of the convolution module} to the last {\it block} of convolution (see Figure \ref{fig:yed-net}), the convolution is less conventional because there is $K^{(l-1)} $ {\it feature maps} ($K^{(l-1)}$ images) as input, denoted $ I^{(l-1)}_k$ with $k=\{1, ..., K^{(l-1)}\}$.
 
The ``convolution'' that will lead to the $k^{th}$ filtered image, $\tilde{I}^{(l)}_k$, resulting from the convolution {\it  block} numbered $l$, is actually the sum of $K^{(l-1)}$ convolutions, such as:

\begin{equation}
\tilde{I}^{(l)}_k = \sum_{i=1}^{i=K^{(l-1)}} I^{(l-1)}_i \star F^{(l)}_{k, i},
\label{eq:I2}
\end{equation}  
with $\{F^{(l)}_{k, i}\}_{i=1}^{i=K^{(l-1)}}$ a set of $K^{(l-1)}$ filters for a given $ k $ value.

This operation is quite unusual since each {\it feature map} is obtained by a {\it sum} of $K^{(l-1)}$ convolutions with a different filter kernel for each convolution. This operation can be seen as a spatial convolution, plus a sum on the channels-axis \footnote{The channels axis is also referred by ``feature maps''-axis, or ``depth''-axis.}. 

This combined operation can be replaced by a separate operation called {\it SeparableConv} or {\it Depthwise Separable Convolutions} \cite{Chollet2017_Xception}, which allows us to integrate a non-linear operation (an activation function) such as a ReLU, between the spatial convolution and the convolution on the ``depth'' axis (for the ``depth'' axis we use a $1\times1$ filter). Thus, the {\it Depthwise Separable Convolution} can roughly be resumed as a weighted sum of convolution which is a more descriptive operation than just a sum of convolution (see Equation \ref{eq:I2}).

If we replace the operation previously described in equation \ref{eq:I2}, by a {\it Depthwise Separable Convolutions} operation integrated within an {\it Inception} module (the Inception allows us to mainly use filters of variable sizes), we obtain a performance improvement \cite{Chollet2017_Xception}. In steganalysis, this has been observed in the article \cite{Zhu-Net2019}, when modifying the first two blocks of the convolution module of Figure \ref{fig:yed-net}.

As a reminder, in this document, we name a {\it convolution block} the set of operations made by one convolution (or many convolutions performed in parallel in the case of an Inception, and/or two convolutions in the case of a Depthwise Separable convolution), a few activation functions, a pooling, and a normalization. These steps can be formally expressed in a simplified way (except in cases with Inception or Depthwise Separable Convolution) in recursive form by linking a {\it feature map} at the input of a block and the {\it feature map} at the output of this block:
\begin{equation}
I^{(l)}_k = norm \left( pool \left( f\left(b^{(l)}_k+\sum_{i=1}^{i=K^{(l-1)}} I^{(l-1)}_i \star F^{(l)}_{k,i}\right) \right) \right),
\label{eq:layer}
\end{equation}
with $b^{(l)}_k \in \mathbb{R}$ the scalar standing for the convolution bias, $f()$ the activation function applied pixel by pixel on the filtered image, $pool()$, the {\it pooling} function that is applied to a local neighborhood, and finally a normalization function.

Note that the kernels of the filters (also called weights) and the bias must be learned and are therefore modified during the back-propagation phase.

\subsection{Activation}
\label{ssec:activation}

\begin{figure*}[ht]
\centering
\FIG{\includegraphics[width=\columnwidth,keepaspectratio=True]{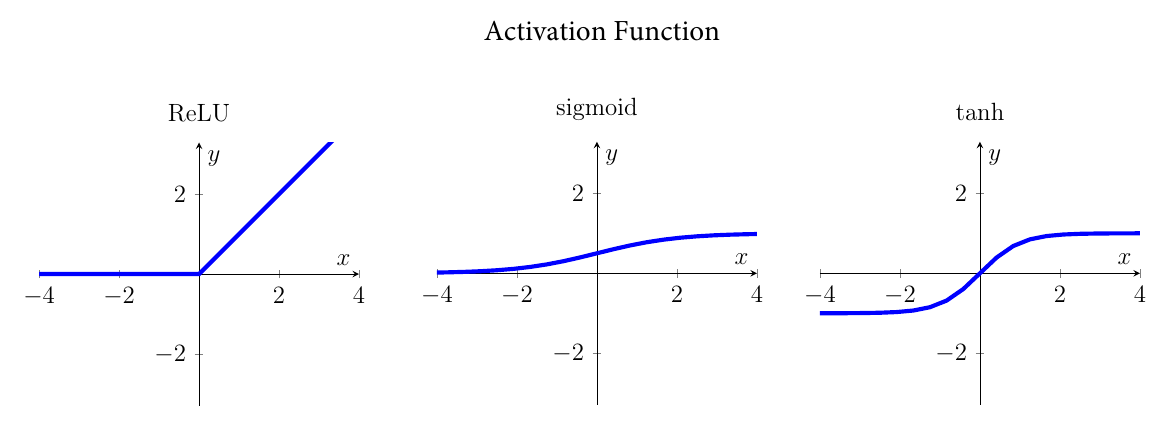}
     \includegraphics[width=\columnwidth,keepaspectratio=True]{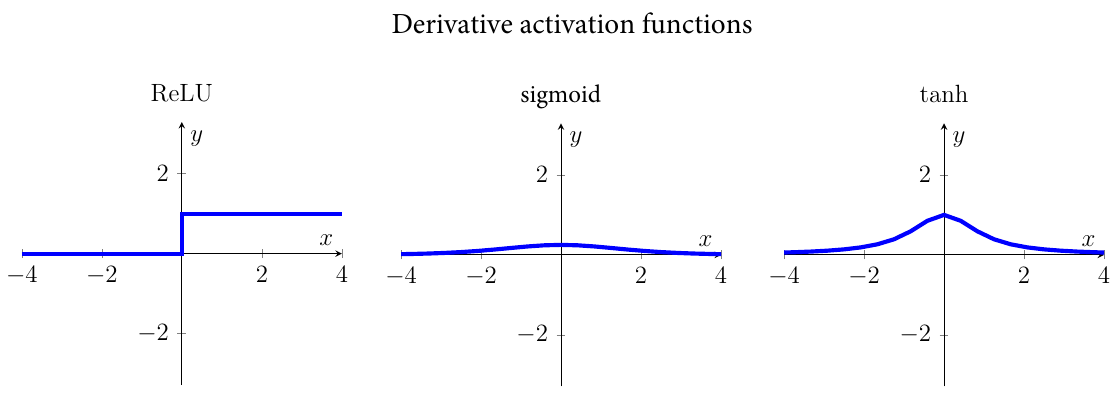}}
    {\caption{Three main activation functions and their derivatives.}}
  \label{fig:activation_function}
\end{figure*}

Once each convolution of a {\it convolution block} has been applied, an {\it activation} function, $f()$ (see Eq. \ref{eq:layer}), is applied on each value of the filtered image, $\tilde{I}^{(l)}_k$ (Eq. \ref{eq:I1} and Eq. \ref{eq:I2}). This function is called the activation function with reference to the notion of binary activation found in the very first work on neuron networks. The activation function can be one of several, for example be an absolute value function $f(x) = |x|$, a sinusoidal function $f(x) = sinus(x)$, a Gaussian function as in \cite{Qian_2015_Deep} $f(x) = \frac{e^{-x^2}}{\sigma^2}$, a ReLU (for {\it Rectified Linear Unit}): $f(x) = max(0,x)$, etc. Figure \ref{fig:activation_function} illustrates some activation functions.

These functions break the linearity resulting from linear filtering performed during convolutions. Non-linearity is a mandatory property that is also exploited in {\it two-step machine-learning approaches}, such as in the ensemble classifier \cite{Kodovsky2012-EnsembleClassifiers} during the weak-classifiers thresholding, or through the final majority vote, or in Rich Models with Min-Max features \cite{Fridrich2012_Rich}. The chosen activation function must be differentiable to perform back-propagation. 

The most often retained solution for the selection of an activation function is one whose derivative requires little calculation to be evaluated. Besides, functions that have low slope regions, such as the hyperbolic tangent, are also avoided, since this type of function can cause the value of the back-propagated gradient to be canceled during back-propagation (the phenomenon of the {\it vanishing gradient}), and thus will make learning impossible. Therefore, in many networks, we very often find the ReLU activation function, or one of its variants. For example, in the Yedroudj-Net network (see figure \ref{fig:yed-net}) we have the absolute value function, the parameterized Hard Tanh function (Trunc function), and the ReLU function. In the SRNet network \cite{Boroumand2018_SRNet} we only find the ReLU function.

\subsection{Pooling}
\label{ssec:pooling}

\begin{figure*}[ht]
\centering
  \FIG{\includegraphics[width=7cm,keepaspectratio=True]{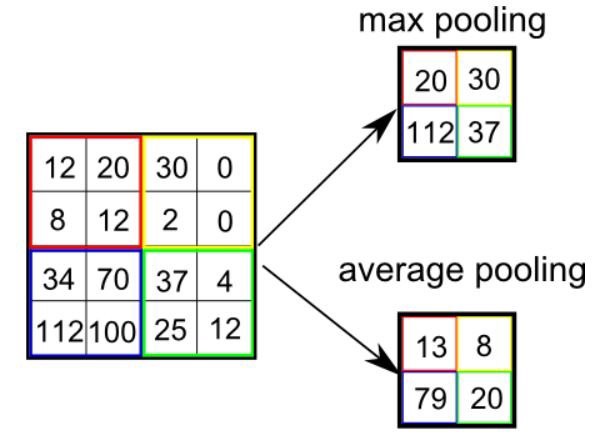}}
      {\caption{Illustration of a maximum pooling and an average pooling.}}
  \label{fig:pool}
\end{figure*}

The pooling operation is used to calculate the {\it average} or the {\it maximum} in a local neighborhood. In the field of classification of objects in images, the maximum pooling guarantees a local invariance in translation when recomputing the features. That said, in most steganalysis networks, it is preferred to use average pooling to preserve stego noise which is of very low power. Figure \ref{fig:pool} illustrates the two pooling operations.

Moreover, pooling is often coupled to a down-sampling operation (when the {\it stride} is greater than 1) to reduce the size (i.e., the height and width) of the resulting {\it feature map} compared to feature maps from the previous block. For example, in Yedroudj-Net (see figure \ref{fig:yed-net}), blocks 2, 3, and 4, reduce by a four-factor the size of the input feature maps. We can consider the pooling operation, accompanied by a stride greater than 1, as conventional sub-sampling with preliminary low-pass filtering. This is useful for reducing the amount of used-memory in the GPU. This step can also be perceived as denoising, and from the point of view of the signal processing, it induces a loss of information. It is probably better not to sub-sample in the first blocks as it was initially highlighted in \cite{Pibre2016}, set up in Xu-Net \cite{Xu2016a}, Ye-Net \cite{Ye2017}, Yedroudj-Net \cite{Yedroudj2018_Net}, and evaluated again in SRNet \cite{Boroumand2018_SRNet}.

\subsection{normalization}
\label{ssec:normalization}

In the first proposed networks in steganalysis, during the period 2014 $-$ {\it beginning of} 2016 (Tan and Li \cite {Tan2014}, Qian {\it et al.} \cite{Qian_2015_Deep}, Pibre {\it and al.} \cite{Pibre2016}), if there was a normalization, it remained local to the spatial neighborhood, with {\it Local Constrast Normalization}, or inter-feature, with the {\it Local Response Normalization}.

A big improvement occurred with the arrival of {\it batch normalisation}. {\it Batch normalization} (BN) was proposed in 2015 \cite{Ioffe2015_BN}, and was widely adopted. This normalization is present in most of the new networks for steganalysis. BN \cite{Ioffe2015_BN} (see Eq. \ref{eq:batch}) consists of normalizing the distribution of each feature of a feature map, so that the average is zero and the variance is unitary, and, if necessary allows re-scaling and re-translating the distribution.

Given a random variable $X$ whose realization is a value $x\in \mathbb{R}$ of the feature map, the BN of this value $x$ is:
\begin{equation}
\label{eq:batch}
BN(x,\gamma,\beta )=\beta +\gamma \frac{x\--E[X]}{\sqrt{Var[X]+\epsilon}},
\end{equation}
with $E[X]$ the expectation, $Var[X]$ the variance, and $\gamma$ and $\beta$ two scalars representing a re-scaling and a re-translation. The expectation $E[X]$ and the variance $Var[X]$ are updated at each batch, while $\gamma$ and $\beta$ are learned by back-propagation. In practice, the BN makes the learning less sensitive to the initialization of parameters \cite{Ioffe2015_BN}, allows us to use a higher learning rate which speeds up the learning process, and improves the accuracy of classification \cite{Chen2017_PNet}.

In Yedroudj-Net, the terms $\gamma$ and $\beta$ are treated by an independent layer called {\it Scale Layer} (See Figure \ref{fig:yed-net}), in the same way as in ResNet \cite{He2016_ResNet}. The increment in performance is very minor.

%%%%%%%%%%%%%%%%%%%%%%%%%%%%%%%%%%%%%%%%%%%%%%%%%%%%%%%%%%%%%%%%%%%%%%%%%%%%%%%%%%%%%%%%%%%%%%%%%%%%%
% SECTION
%%%%%%%%%%%%%%%%%%%%%%%%%%%%%%%%%%%%%%%%%%%%%%%%%%%%%%%%%%%%%%%%%%%%%%%%%%%%%%%%%%%%%%%%%%%%%%%%%%%%%
\section{Memory / time complexity and efficiency}
\label{sec:complexity}

Learning a network can be considered as the optimization of a function with many unknown parameters, thanks to the use of a well-thought out stochastic gradient descent. In the same way as traditional neural networks, the CNNs used for steganalysis have a large number of parameters to learn. As an example, without taking into account Batch Normalization and Scale parameters, the Xu-Net \cite {Xu2016a} network described in the paper \cite{Yedroudj2018_Net} has a number of parameters in the order of 50,000. In comparison, the network Yedroudj-Net \cite{Yedroudj2018_Net}, has a number of unknown parameters in the order of 500,000.

In practice, using a previous-generation GPU (Nvidia TitanX) on an Intel Core i7-5930K at 3.50 GHz $\times$ 12 with 32 GB of RAM, it takes less than a day to learn the Yedroudj-Net network using 4,000 pairs of $256 \times 256$ cover/stego images of the `` BOSS'' \cite{Bas2011-BOSS}, three days on 14,000 pairs of $256 \times 256$ cover/stego images of ``BOSS +  BOWS2'' \cite{BOWS2008}, and more than seven days on the 112,000 pairs of $256 \times 256 $ cover/stego images of ``BOSS + BOWS2 + a virtual database augmentation'' \cite{Yedroudj2018_DatabaseAugmentation}. These long learning times are because the databases are large and have to be browsed repeatedly, so that the back-propagation process makes converge the network.

Due to the large number of parameters to be learned, neural networks need a database containing a large number of examples to be in {\it the power-law region} \cite {Hestness2017} allowing comparisons between different networks. In addition, the examples within the learning database must be sufficiently diversified to obtain a good generalization of the network. For CNN steganalysis, with current networks (in 2018), the number of examples needed to reach a region of {\it good performance} (that is, as good as using a Rich Model\cite{Fridrich2012_Rich} with an Ensemble Classifier \cite{Kodovsky2012-EnsembleClassifiers}), in the case where there is no cover-source mismatch, is most likely in the order of 10,000 images (5,000 covers and 5,000 stegos) when the size is $256 \times 256$ pixels \cite{Yedroudj2018_DatabaseAugmentation}. However, the number of examples is still insufficient \cite{Yedroudj2018_DatabaseAugmentation} in the sense that performance can be increased simply by increasing the number of examples. The so-called {\it irreducible error region} \cite {Hestness2017} probably requires more than a million images \cite{Zeng2017_Millions}; Therefore, there should be at least 100 times more images for the learning phase. In addition to this, it is necessary to be able to work with larger images. It is therefore evident that in the future it will be essential to find one or more solutions to reach the region of {\it irreducible error}. This can be done with huge databases, and several weeks or months of apprenticeships, or by transfer learning, by using better networks, or with solutions yet to be conceived.

Note that of course, there are recommendations to increase performance and it may be possible to achieve the {\it irreducible error} region faster. We can use transfer learning \cite {Qian2016_Transfer} and/or curriculum learning \cite{Ye2017} to start learning from a network that has already learned. We can use a set of CNNs \cite{Xu2016b}, or a network made of sub-networks \cite{Li2018_ReST-Net}, which can save a few percentage points on accuracy. We can virtually increase the database \cite{Krizhevsky2012_Convnet}, but this does not solve the problem of increasing the learning time. We can add images of a database that is similar to the test database, for example when BOSS and BOWS2 are used for learning, in the case where the test is realized on BOSS \cite{Ye2017}, \cite{Yedroudj2018_DatabaseAugmentation}. It is nevertheless not obvious that in practice we can have access to a database similar to the database to be tested. We can (i) predict the acquisition devices that produced the images of the test database, then (ii) make new acquisitions with these devices (to be purchased), and (iii) finally perform images development similar to the one used to generate the test database, and all this in order to increase the learning database \cite{Yedroudj2018_DatabaseAugmentation}. Again, this approach is difficult to implement and time-consuming.

Note that a general rule shared by people playing with Kaggle competitions is that the main practical rules to win are~\cite{Kuzin2018_2KaggleCameraId}~\footnote{The authors of \cite{Kuzin2018_2KaggleCameraId} finished second at the Kaggle competition for {\it IEEE's Signal Processing Society - Camera Model Identification - Identify from which camera an image was taken}. \\{\scriptsize \url{https://www.kaggle.com/c/sp-society-camera-model-identification}}. \\{\scriptsize \href{https://towardsdatascience.com/forensic-deep-learning-kaggle-camera-model-identification-challenge-f6a3892561bd}{https://towardsdatascience.com/forensic-deep-learning-kaggle-camera-model-identification-challenge-f6a3892561bd}}.}: (i) to use an ensemble of modern networks (ResNet, DenseNet, etc.) that have learned for example on ImageNet, and then use transfer learning, and (ii) to do data-augmentation, (iii) to eventually collect data to increase the learning database size.

%%%%%%%%%%%%%%%%%%%%%%%%%%%%%%%%%%%%%%%%%%%%%%%%%%%%%%%%%%%%%%%%%%%%%%%%%%%%%%%%%%%%%%%%%%%%%%%%%%%%%
% SECTION
%%%%%%%%%%%%%%%%%%%%%%%%%%%%%%%%%%%%%%%%%%%%%%%%%%%%%%%%%%%%%%%%%%%%%%%%%%%%%%%%%%%%%%%%%%%%%%%%%%%%%
\section{Link between Deep-learning and past approaches}
\label{sec:linktopast}

In previous Sections, we explained that deep-learning consisted of minimizing a function with many unknown parameters with a technique similar to gradient descent. In this subsection, we establish links with previous research on the subject in the steganography/steganalysis community. This sub-section tries to make links with past research in this domain and is an attempt to demystify deep learning.

Convolution is an essential part of CNN networks. Learning filter kernels (weights) are carried out by minimizing the classification error using the back-propagation procedure. It is, therefore, a simple optimization of filter kernels. Such a strategy can be found as early as 2012 in a two-step approach using Rich Models and an Ensemble Classifier in the article \cite{Holub2012_Filters}. The kernel values used to calculate the feature vector are obtained by optimization via the simplex algorithm. In this article, the goal is to minimize the probability of classification error given by an Ensemble Classifier in the same way as with a CNN. CNNs share the same goal of building custom kernels that are well suited to steganalysis.

Looking at the first {\it block} of convolution just after the pre-processing {\it block} (Ye-Net \cite{Ye2017}, Yedroudj-Net \cite{Yedroudj2018_Net}, ReST-Net \cite{Li2018_ReST-Net}, etc.), the convolutions act as a multi-band filtering performed on the residuals obtained from the pre-processing block (see Figure \ref{fig:yed-net}). For this first block, the network analyzes the signal residue in different frequency bands. In the past, when computing Rich Models \cite{Fridrich2012_Rich}, some approaches have applied a similar idea thanks to the use of a filter bank. Some approaches make a spatio-frequency decomposition via the use of Gabor filters (GFR Rich Models) \cite{Song2015_RichJPEGGabor}, \cite {Xia2017_GFR}, some use Discrete Cosinus filters (DCTR Rich Models) \cite{Holub2015_DCTR}, some use Steerable Gaussian filters \cite{Abdulrahman2016-SteerableGaussian}, and some make a projection on random carriers (PSRM Rich Models) \cite{Holub2013_PSRM}, etc. For all these Rich Models, the results of the filtering process is then used to calculate a histogram (co-occurrence matrix) which is in turn used as a vector of features. The first convolution block of CNNs for steganalysis thus share similarities with the spatio-frequency decomposition of some Rich Models.

From the convolution blocks that start to down-sample the feature maps, there is a summation of the results of several different convolutions. This amounts to accumulating signs of the presence of a signal (the stego noise) by observing clues in several bands. We do not find such a principle in previous research. The only way to accumulate evidence was based on the computation of a histogram \cite{Fridrich2012_Rich, Holub2013_PSRM}, but this approach is different from what is done in CNNs. Note that in the article \cite{Sedighi2017_HistoCNN}, the authors explore how to incorporate the histogram computation mechanism into a CNN network, but the results are not encouraging. Thus, starting from the second block, the mechanism involved to create a latent space separating the two classes, i.e. to obtain a feature vector per image, which makes it possible to distinguish the covers from the stegos, is different from that used in Rich Models. Similarly, some past techniques such as non-uniform quantization \cite{Pevny2012_NonUniformQuantization}, features selection \cite{Chaumont2012-EC-FS}, dimension reduction \cite{Pevny2013_ReductionDimensionCLS}, are not directly visible within a CNN.

A brick present in most convolution blocks is the normalization of feature maps. Normalization has often been used in steganalysis, for example in \cite{Kouider2013}, \cite{Cogranne2014_EC_HypothesisTest}, \cite{Boroumand2017_Normalization}, etc. Within a CNN, normalization is performed among other things to obtain comparable output values in each feature map.

The activation function introduces a non-linearity in the signal and thus makes it possible to have many convolution blocks. This non-linearity is found in the past, for exemple in the Ensemble Classifier through the majority vote \cite {Kodovsky2012-EnsembleClassifiers}, or in Rich Models with the Min or Max operations \cite{Fridrich2012_Rich}.

The structure of a CNN network and the components that improve the performance of a network are now better understood in practice. As we saw previously, there is in a CNN, some parts that are similar to propositions made in steganalysis in the past. Some elements of a CNN are also explained by the fact that they are guided by computational constraints (uses of simple differentiable activation function like ReLU), or that they facilitates the convergence (non-linearity allows convergence, activation function should not be too flat or steep, in order to avoid vanishing gradient or rapid variation, the shortcut allows us to avoid vanishing gradient during back-propagation, and thus allows us to create deeper networks, the batch normalization, the initialization such as Xavier, the optimization such as Adam, etc). Note that some of the ideas present in CNNs also come from the theory of optimization of differentiable functions.

Although it is easy to use a network in practice, and to have some intuition about its behavior, it still lacks theoretical justification. For example, what is the right number of parameters according to the problem? In the coming years, there is no doubt that the building of a CNN network adapted for steganalysis could go through an automatic adjustment of its topology, in this spirit, the work on AutoML and Progressive Neural Architecture Search (PNAS) \cite{Liu2018_PNAS}, \cite{Pham2018_ENAS} are of interest. That said the theory must also try to explain what is happening inside the network. One can notably look at the work of St\'{e}phane Mallat \cite{Mallat2016} for an attempt to explain a CNN from a signal processing point of view. Machine learning theorists can also better explain what happens in a network and why this mathematical construction works so well.

To conclude this discussion on the links between two-step learning approaches and deep learning approaches, CNN as well as two-step (Rich Models $+$ Ensemble Classifier) approaches are not able to cope with cover-source mismatch \cite{Cancelli2008, Fridrich2011-HugoProcessDiscovery}. This is a defect used by detractors \footnote{See Gary Marcus' web-press article \url{https://medium.com/@GaryMarcus/the-deepest-problem-with-deep-learning-91c5991f5695}.}
of neural network approaches in domains such as object recognition \cite{Alcorn2018_arXiv}. CNNs learn a distribution, but if it differs in test phase, then the network cannot detect it. Maybe the ultimate goal is for the network to ``understand'' that the test database is not distributed as the learning database?

%%%%%%%%%%%%%%%%%%%%%%%%%%%%%%%%%%%%%%%%%%%%%%%%%%%%%%%%%%%%%%%%%%%%%%%%%%%%%%%%%%%%%%%%%%%%%%%%%%%%%
% SECTION
%%%%%%%%%%%%%%%%%%%%%%%%%%%%%%%%%%%%%%%%%%%%%%%%%%%%%%%%%%%%%%%%%%%%%%%%%%%%%%%%%%%%%%%%%%%%%%%%%%%%%
\section{The different networks used over the period 2015-2018}
\label{sec:periode15-18}

\begin{figure*}[tb]
  \FIG{\includegraphics[width=\columnwidth]{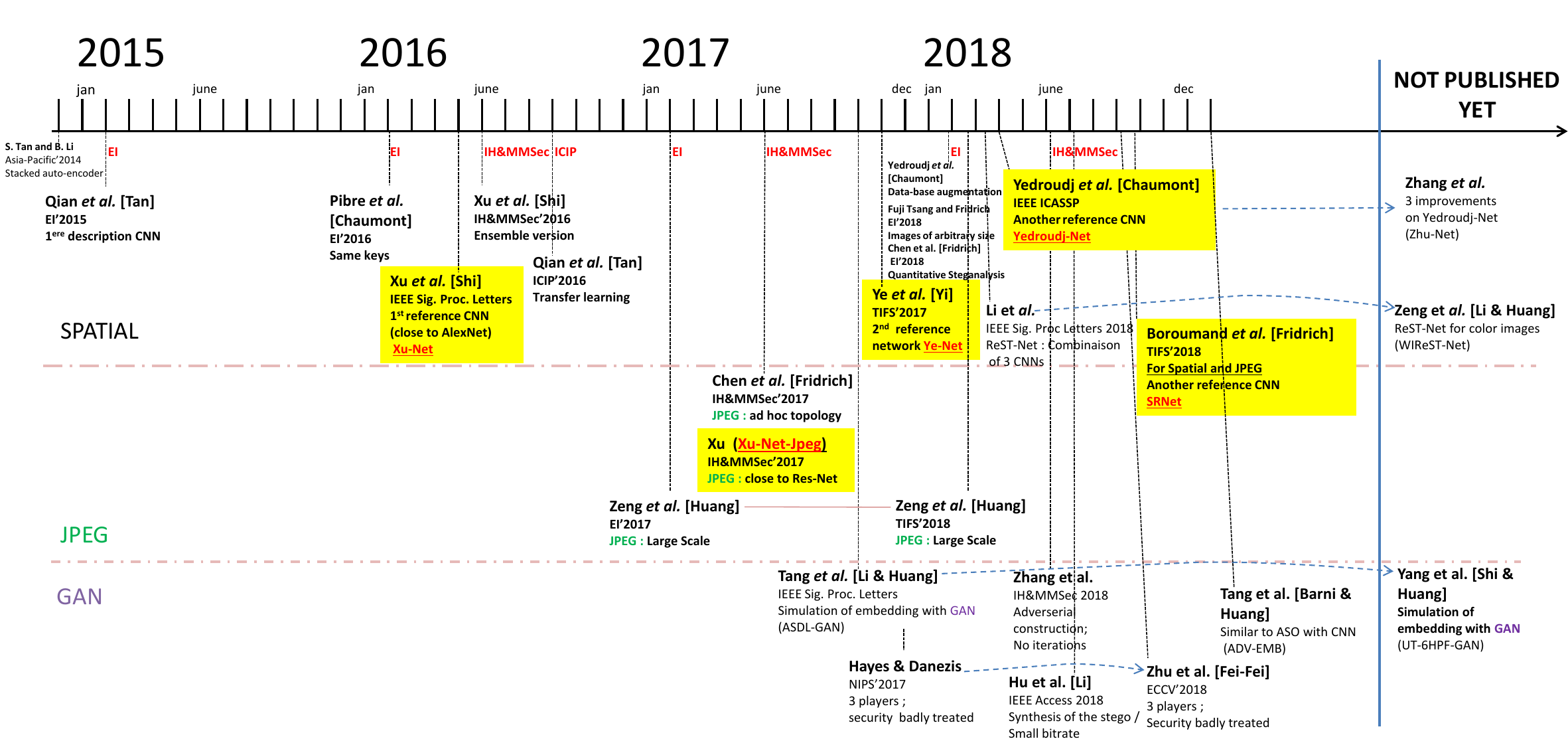}}
		  %{\caption{Chronology from 2015 to 2018.}}
			{\caption{{\scriptsize Chronology of the main CNNs for steganography and steganalysis from 2015 to 2018}.}}
  \label{fig:chronology}
\end{figure*}

A chronology of the main CNNs proposed for steganography and steganalysis from 2015 to 2018 are given in Figure \ref{fig:chronology}. The first attempt to use Deep Learning methods for steganalysis date back to the end of 2014 \cite{Tan2014} with auto-encoders. At the beginning of 2015, Qian {\it et al.} \cite{Qian_2015_Deep} proposed to use Convolutional Neural Networks. One year later Pibre {\it et al.} \cite{Pibre2016} proposed to pursue the study. 

\begin{figure*}[ht]
\centering
  \FIG{\includegraphics[width=\columnwidth,keepaspectratio=False]{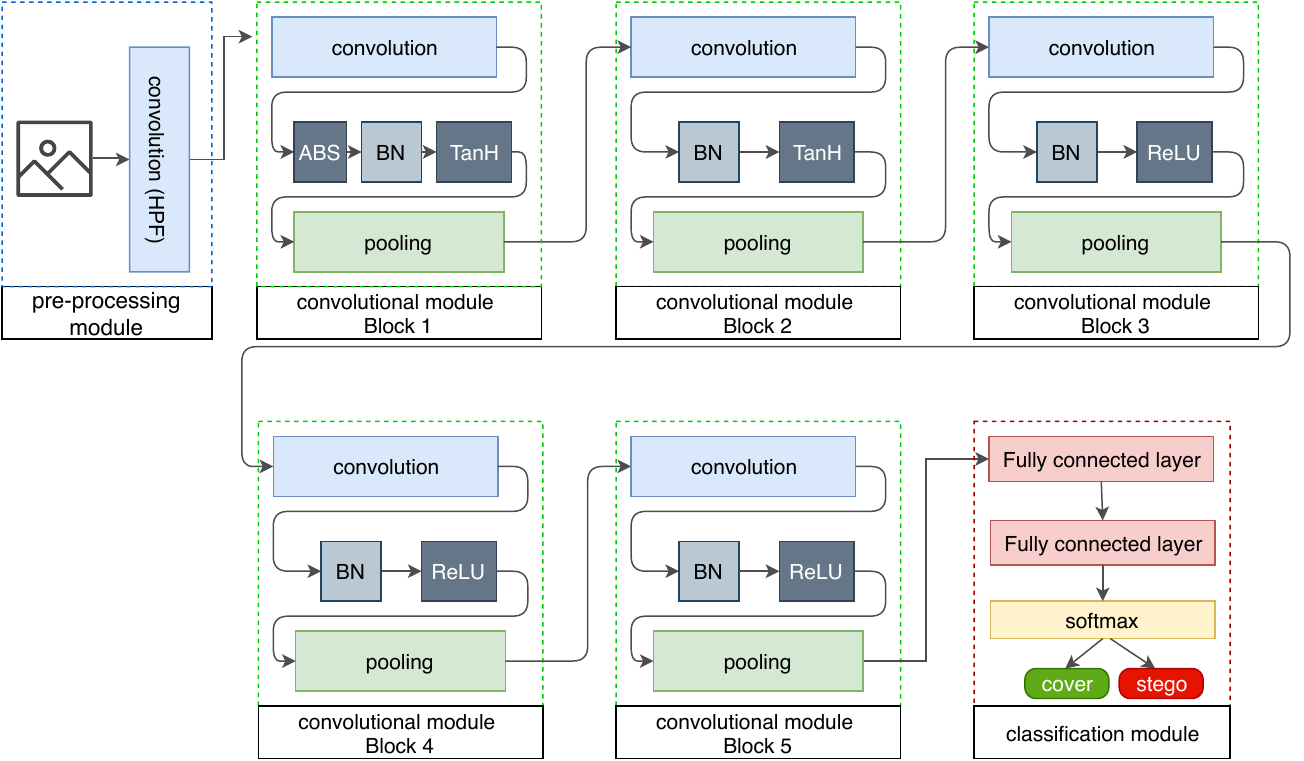}}
      {\caption{Xu-Net overall architecture.}}
  \label{fig:Xu-Net}
\end{figure*}

In 2016, the first results, close to those of current state-of-the-art methods (Ensemble Classifier + Rich Models), were obtained with an ensemble of CNNs \cite{Xu2016b}; See Figure \ref{fig:Xu-Net}. The Xu-Net \footnote{In this chapter, we reference {\it Xu-Net} a CNN similar to the one given in \cite{Xu2016a}, and not to the ensemble version \cite{Xu2016b}.} \cite{Xu2016a} CNN is used as a {\it base learner} of an ensemble of CNNs. 

\begin{figure*}[ht]
\centering
  \FIG{\includegraphics[width=7cm,keepaspectratio=False]{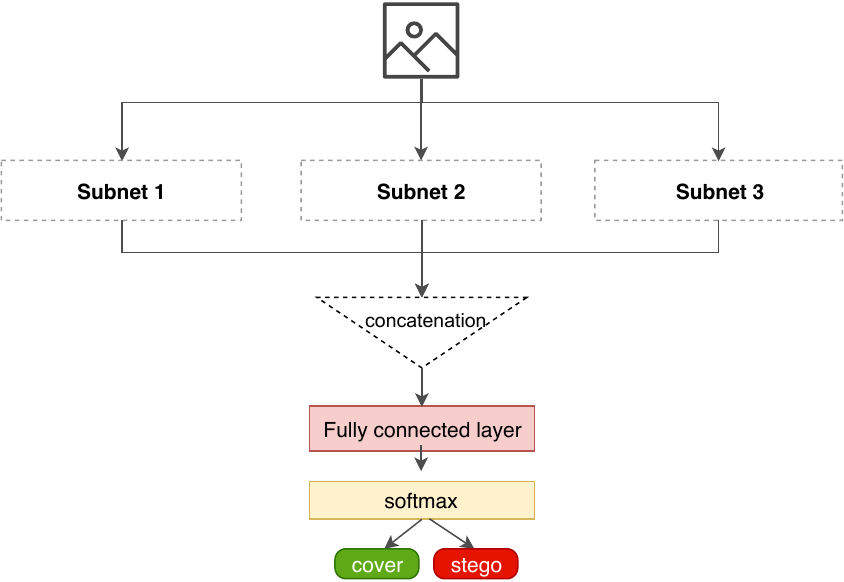}}
      {\caption{ReST-Net overall architecture.}}
  \label{fig:ReST-Net}
\end{figure*}

\begin{figure*}[ht]
\centering
  \FIG{\includegraphics[width=\columnwidth,keepaspectratio=False]{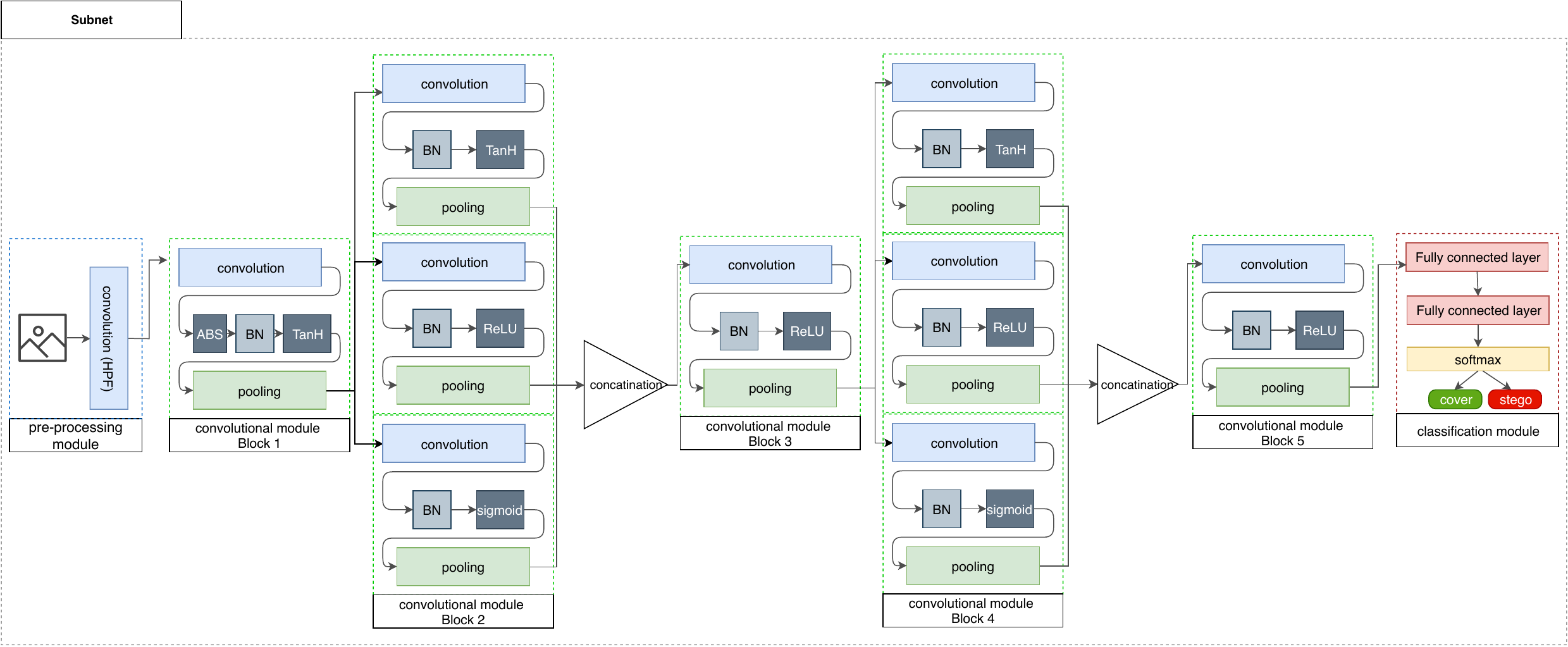}}
      {\caption{ReST-Net sub-network.}}
  \label{fig:ReST-Net-sub}
\end{figure*}

Other networks were proposed in 2017, this time for JPEG steganalysis. In \cite{Zeng2017} \cite{Zeng2017_Millions} (See Figures \ref{fig:ReST-Net} and \ref{fig:ReST-Net-sub}), authors proposed a pre-processing inspired by Rich Models, and the use of a large learning database. The results were close to those of existing state-of-the-art methods (Ensemble Classifier + Rich Models). In \cite{Chen2017_PNet}, the network is built with a {\it phase-split} inspired by the JPEG compression process. An ensemble of CNNs was required to obtain results that were slightly better than those obtained by current best approach. In Xu-Net-Jpeg \cite{Xu2017}, a CNN inspired by ResNet \cite{He2016_ResNet} with the {\it shortcut connection} trick, and 20 blocks also improved the results in terms of accuracy. Note that in 2018 the ResDet \cite{Huang2018_ResDet} proposed a variant of {\bf Xu-Net-Jpeg} \cite{Xu2017} with similar results.


These results were highly encouraging, but regarding the gain obtained in other image processing tasks using Deep Learning methods \cite{Lecun2015}, the steganalysis results represented less than a 10\% improvement compared to the classical approaches that use an Ensemble Classifier \cite{Kodovsky2012-EnsembleClassifiers} with Rich Models \cite{Fridrich2012_Rich}, \cite{Xia2017} or Rich Models with a Selection-Channel Awareness \cite{Denemark2014_maxSRM}, \cite{Denemark2016_SCA_Sigma}, \cite{Denemark2016_SCA_Delta}. The revolutionary significant gain in the use of deep learning, observed in other areas of signal processing, was not yet present for steganalysis. In 2017, the main trends to improve CNN results were using an ensemble of CNNs, modifying the topology by mimicking Rich Models extraction process or using ResNet. In most of the cases, the design or the experimental effort was very high for a very limited improvement of performance in comparison to networks such as AlexNet \cite{Krizhevsky2012_Convnet}, VGG16 \cite{Simonyan2015_VGG16}, GoogleNet \cite{Szegedy2015_GoogleNet}, ResNet \cite{He2016_ResNet}, etc, that inspired this research.

\begin{figure*}[ht]
\centering
   \FIG{\includegraphics[width=\columnwidth,keepaspectratio=False]{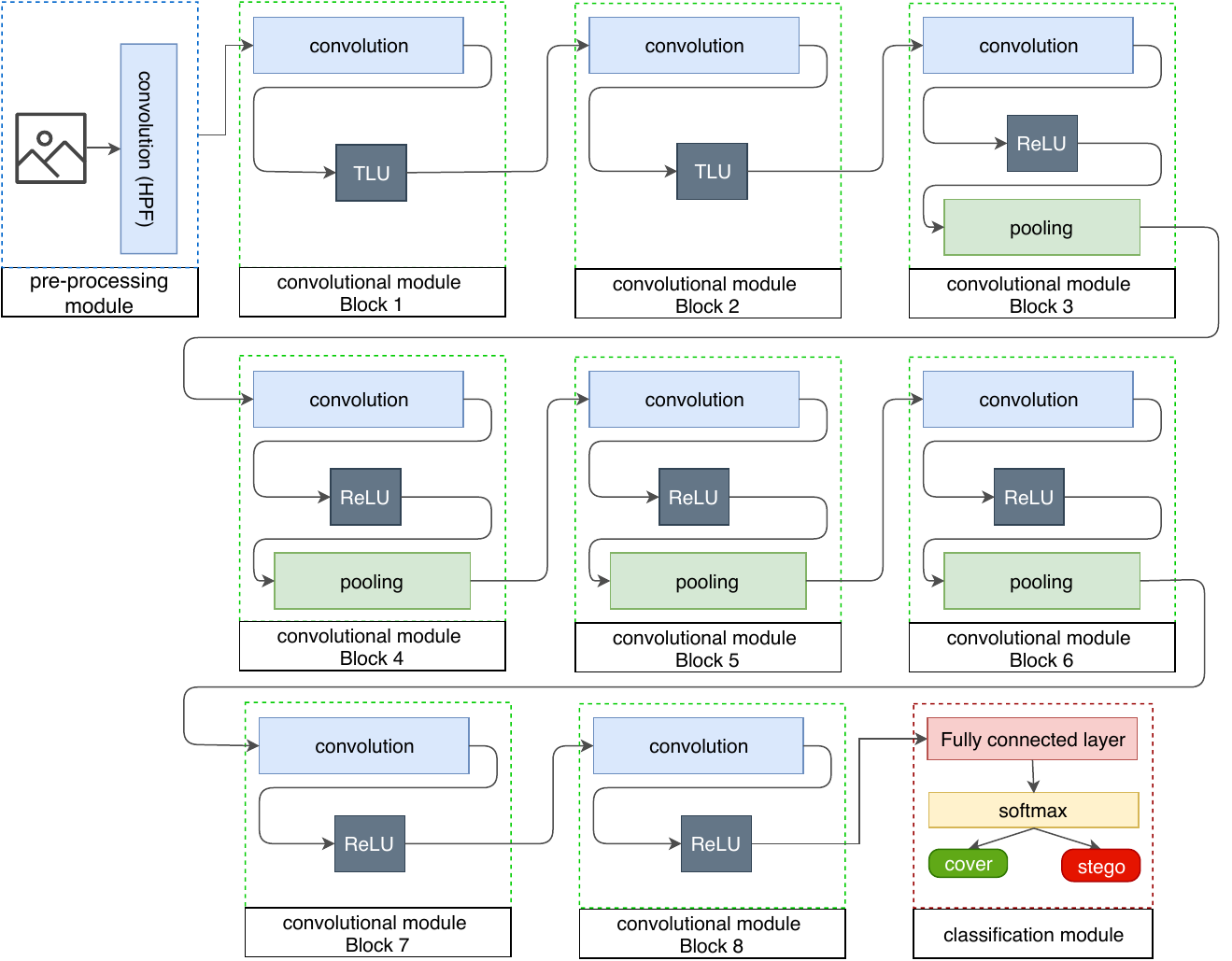}}
       {\caption{Ye-Net overall architecture.}}
  \label{fig:Ye-Net}
\end{figure*}

\begin{figure*}[!htb]
\centering
  \FIG{\includegraphics[width=\columnwidth]{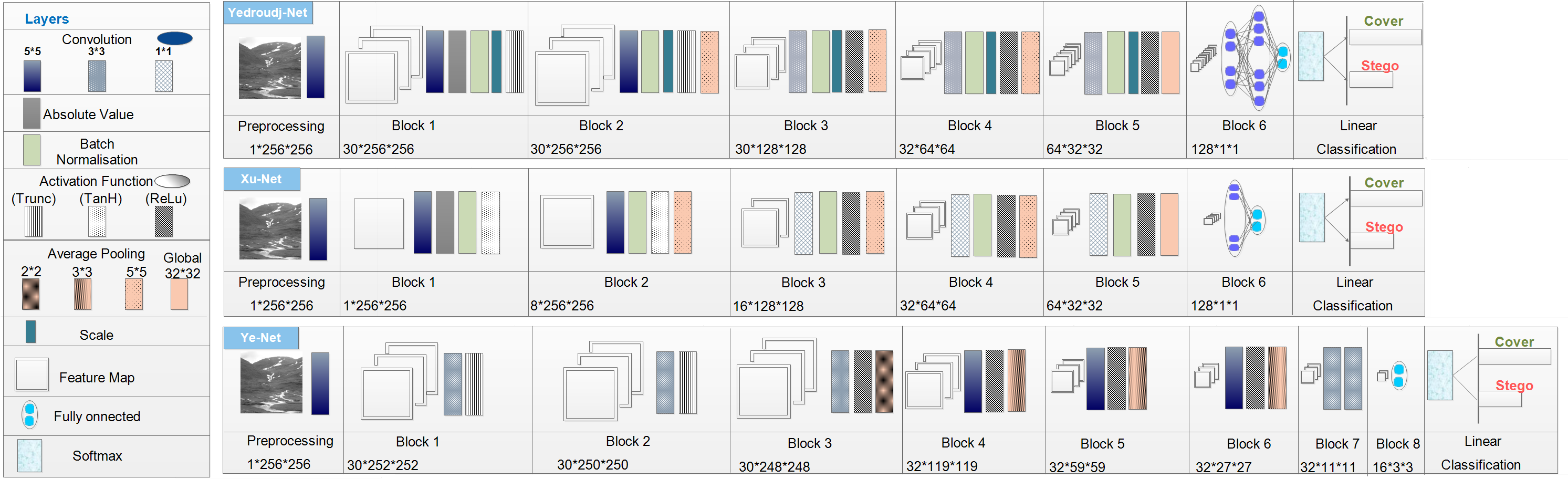}}
      {\caption{Comparison of Yedroudj-Net, Xu-Net, and Ye-Net architectures.}}
\label{fig:nets}
\end{figure*}

\begin{figure*}[!htb]
\centering
  \FIG{\includegraphics[width=\columnwidth]{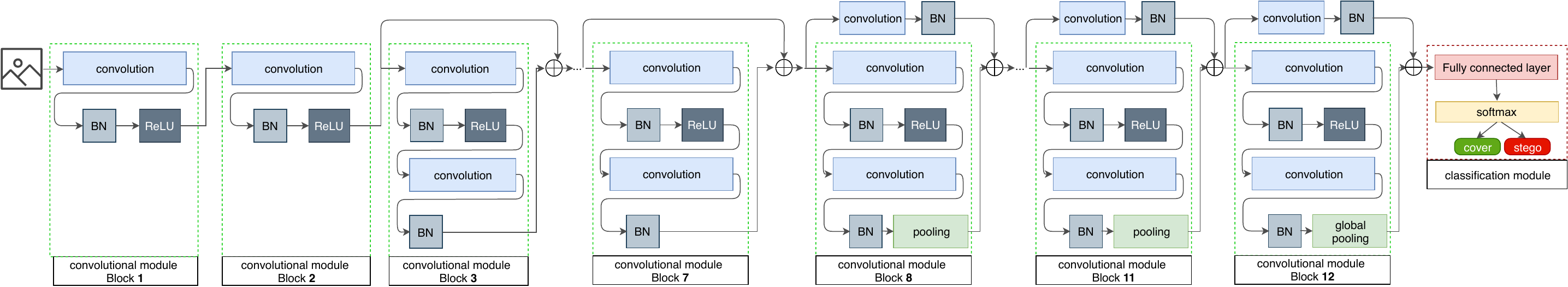}}
      {\caption{SRNet network.}}
\label{fig:SRNet}
\end{figure*}

By the end of 2017 and early 2018, the studies had strongly concentrated on spatial steganalysis. Ye-Net \cite{Ye2017} (See Figure \ref{fig:Ye-Net}), Yedroudj-Net\footnote{Yedroudj-Net source code: {\tiny\url{https://github.com/yedmed/steganalysis_with_CNN_Yedroudj-Net}}.} \cite{Yedroudj2018_DatabaseAugmentation, Yedroudj2018_Net} (See Figure \ref{fig:nets}), ReST-Net \cite{Li2018_ReST-Net} (See Figures \ref{fig:ReST-Net} and \ref{fig:ReST-Net-sub}), SRNet\footnote{SRNet source code: {\tiny\url{https://github.com/Steganalysis-CNN/residual-steganalysis}}.} \cite{Boroumand2018_SRNet} (See Figures \ref{fig:SRNet}) have been published respectively in November 2017, January 2018, May 2018, and May 2019 (with an online version in September 2018). All these networks clearly surpass the ``old'' two-step machine learning paradigm that was using an Ensemble Classifier \cite{Kodovsky2012-EnsembleClassifiers} and Rich Models \cite{Fridrich2012_Rich}. Most of these networks can learn with a modest database size (i.e. around 15,000 pairs cover/stego of 8-bits-coded images of $256\times256$ pixels size from BOSS+BOWS2). 

In 2018, the best networks were Yedroudj-Net \cite{Yedroudj2018_Net}, ReST-Net \cite{Li2018_ReST-Net}, and SRNet \cite{Boroumand2018_SRNet}. Yedroudj-Net is a small network that can learn on a very small database and can be effective even without using the tricks known to improve performance such as transfer learning \cite{Qian2016_Transfer} or virtual augmentation of the database \cite{Ye2017}, etc. This network is a good candidate when working on GANs. It is better than Ye-Net \cite{Ye2017}, and can be improved to face other more recent networks \cite{Zhu-Net2019}. ReST-Net \cite{Li2018_ReST-Net} is a huge network made of three sub-networks which uses various pre-processing filter banks. SRNet \cite{Boroumand2018_SRNet} is a network that can be adapted to spatial or Jpeg steganalysis. It requires various tricks such as virtual augmentation and transfer learning, and therefore requires a bigger database compared to Yedroudj-Net. These three networks are described in Section \ref{ssec:nsca}.

To resume, from 2015 to 2016, publications were in spatial steganalysis, in 2017, the publications were mainly on JPEG steganalysis. In 2018, publications were again mainly concentrated on spatial steganalysis. Finally, at the end of 2017, the first publications using GANs appeared. In Section \ref{sec:GAN} we present the new propositions using steganography by deep-learning, and give classification per family.

In the next subsection, we report on the most successful networks until the end of 2018, for various scenarios. In Section \ref{ssec:nsca}, we describe the {\it Not-Side-Channel-Aware} (Not-SCA) scenario, in Section \ref{ssec:sca} we discuss the scenario known as {\it Side-Channel-Aware} (SCA), in Sections \ref{ssec:jpeg_nsca} we deal with JPEG steganalysis {\it Not-SCA} and {\it SCA} scenarios. In Section \ref{ssec:mismatch} we very briefly discuss cover-source mismatch, although for the moment the proposals using a CNN do not exist.

We will not tackle the scenario of CNN invariant to the size of the images because it is not yet mature enough. This scenario is briefly discussed in Section \ref{ssec:classification}, and the papers of Yedroudj-Net \cite{Yedroudj2018_Net}, Zhu-Net \cite{Zhu-Net2019}, or Tsang {\it et al.} \cite{Tsang2018_ArbitrarySize}, give firsts solutions.

We will not approach the scenario of quantitative steganalysis per CNN, which consists in estimating the embedded payload size. This scenario is very well examined in the paper \cite{Chen2018_Quantitative} and serves as a new state-of-the-art method. The approach surpasses the previous state-of-the-art approaches \cite{Kodovsky2013_Quantitative} \cite{Zakaria2018_Comparative} that rely on Rich Models, an Ensemble of trees, and an efficient normalization of features.

Nor will we discuss batch steganography and pooled steganalysis with CNNs which has not yet been addressed, although the work presented in \cite{Zakaria2019_Pooled} using two-stage machine learning can be extended to deep learning.

\subsection{The spatial steganalysis Not-Side-Channel-Aware (Not-SCA)}
\label{ssec:nsca}

In early 2018 the most successful spatial steganalysis approach is the Yedroudj-Net \cite{Yedroudj2018_Net} method (See Figure \ref{fig:Ye-Net}). The experiments comparisons were carried out on the BOSS database which contains 10,000 images sub-sampled to $256\times256$ pixels. For a fair comparison, the experiments were performed by comparing the approach to Xu-Net without Ensemble \cite{Xu2016a}, to the Ye-Net network in its Not-SCA version \cite{Ye2017}, and also to Ensemble Classifier \cite{Kodovsky2012-EnsembleClassifiers} fed by Spatial-Rich-Models \cite{Fridrich2012_Rich}. Note that Zhu-Net \cite{Zhu-Net2019} (not yet published when writing this chapter) offers three improvements to Yedroudj-Net that allows it to be even more efficient. The improvements reported by Zhu-Net \cite{Zhu-Net2019} are the update to the kernel filters of the pre-processing module (in the same vein as what has been proposed by Matthew Stamm's team in Forensics \cite{Bayar2016_HPFilters}), replacing the first two convolution blocks with two modules of {\it Depthwise Separable Convolutions} as proposed in \cite{Chollet2017_Xception}, and finally replacing global average pooling with a {\it Spatial Pyramid Pooling (SPP)} module as in \cite{He2014_SPP}.

In May 2018 the ReST-Net \cite{Li2018_ReST-Net} approach was proposed (See Figures \ref{fig:ReST-Net} and \ref{fig:ReST-Net-sub}). It consists of agglomerating three networks to form a {\it super-network}. Each sub-net is a modified Xu-Net like network \cite{Xu2016a} resembling the Yedroudj-Net \cite{Yedroudj2018_Net} network, with an Inception module on block 2 and block 4. This Inception module contains filters of the same size, with a different activation function for each ``path'' (TanH, ReLU, Sigmoid). The first subnet performs pre-processing with 16 Gabor filters, the second sub-network pre-processing with 16 SRM linear filters, and the third network pre-processing with 14 non-linear residuals (min and max calculated on SRM). The learning process requires four steps (one step per subnet and then one step for the {\it super-network}). The results are 2-5\% better than Xu-Net for S-UNIWARD \cite{Holub2014_UNIWARD}, HILL \cite{Li2014_HILL}, CMD-HILL \cite{Li2015_HILL-CMD} on the BOSSBase v1.01 \cite{Bas2011-BOSS} 512 $\times$ 512. Looking at the results, it is the concept of Ensemble that improves the performances. Taken separately, each sub-net has a lower performance. At the moment, no comparison in a fair framework was made between an Ensemble of Yedroudj-Net and ReST-Net.

In September 2018 the SRNet \cite{Boroumand2018_SRNet} approach became available online (See Figures \ref{fig:SRNet}). It proposes a deeper network than previous networks, which is composed of 12 convolution blocks. The network does not perform pre-processing (the filters are learned) and sub-samples the signal only from the 8th convolution block. To avoid the problem of vanishing gradient, blocks 2 to 11 use the shortcut mechanism. The Inception mechanism is also implemented from block 8 during the pooling (sub-sampling) phase. The learning database is augmented with the BOWS2 database as in \cite{Ye2017} or \cite{Yedroudj2018_DatabaseAugmentation}, and a curriculum training mechanism \cite{Ye2017} is used to change from a standard payload size of 0.4 bpp to other payload sizes. Finally, gradient descent is performed by Adamax \cite{Kingma2015_Adam}. The network can be used for spatial steganalysis (Not-SCA), for informed (SCA) spatial steganalysis (see Section \ref{ssec:sca}) and for JPEG steganalysis (see Section \ref{ssec:jpeg_nsca} Not-SCA or SCA). Overall the philosophy remains similar to previous networks, with three parts: pre-processing (with learned filters), convolution blocks, and classification blocks. With a simplified vision, the network corresponds to the addition of 5 blocks of convolution without pooling, just after the first convolution block of Yedroudj-Net network. To be able to use this large number of blocks on a modern GPU, authors must reduce the number of feature maps to 16, and in order to avoid the problem of vanishing gradients, they must use the trick of residual shortcut within the blocks as proposed in \cite{He2016_ResNet}. Note that preserving the size of the signal in the first seven blocks is a radical approach. This idea has been put forward in \cite{Pibre2016} where the suppression of pooling had clearly improved the results. The use of modern brick like shortcuts or Inception modules also enhances performance.

It should also be noted that the training is completed end-to-end without particular initialization (except when there is a curriculum training mechanism). In the initial publication \cite{Boroumand2018_SRNet}, SRNet network was not compared to Yedroudj-Net \cite{Yedroudj2018_Net}, or to Zhu-Net \cite{Zhu-Net2019}, but later, in 2019, in \cite{Zhu-Net2019} all these networks have been compared and the update of Yedroudj-Net i.e. Zhu-Net gives performances of 1\% to 4\% improvement over SRNet, and 4\% to 9\% improvement over Yedroudj-Net, when using the usual comparison protocol. Note that Zhu-Net is also better than the network {\it Cov-Pool} published at IH\&MMSec'2019 \cite{Deng2019_Cov-Pool}, and whose performances are similar to SRNet.

\subsection{The spatial steganalysis Side-Channel-Informed (SCA)}
\label{ssec:sca}

At the end of 2018, two approaches combined the knowledge of the selection channel, the SCA-Ye-Net (which is the SCA version of Ye-Net) \cite{Ye2017} and the SCA-SRNet (which is the SCA version of SRNet) \cite{Boroumand2018_SRNet}. The idea is to use a network which is used for non-informed steganalysis and to inject not only the image to be steganalyzed, but also the modification probability map. It is thus assumed that Eve knows, or can have a good estimation \cite{Sedighi2015_ImpreciseSCA} of the modification probability map, i.e. Eve has access to side-channel information. 

\begin{figure*}[ht]
\centering
   \FIG{\includegraphics[width=7cm,keepaspectratio=False]{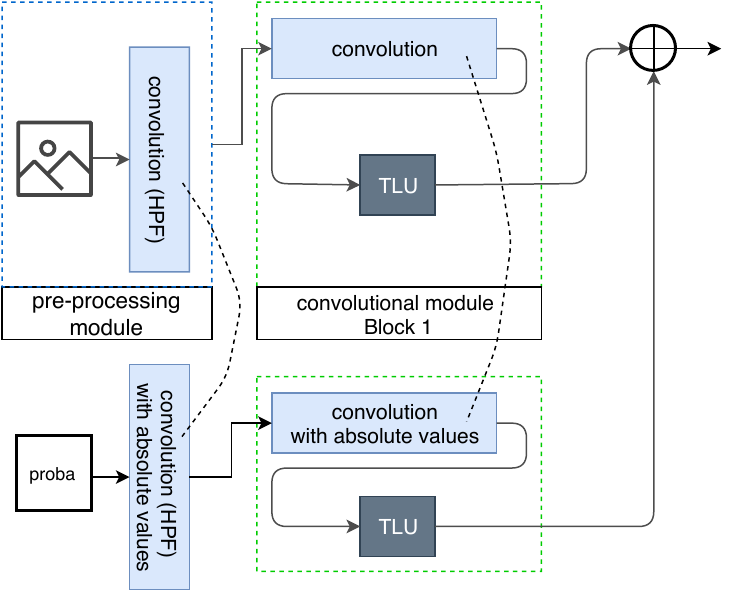}}
       {\caption{Integration of the modification probability map in a CNN.}}
  \label{fig:SCA-spatial}
\end{figure*}

The modification probability map is given to the pre-processing block SCA-Ye-Net \cite{Ye2017}, and equivalently to the first convolution block for SCA-SRNet \cite{Boroumand2018_SRNet}, but the kernel values are replaced by their absolute values. After the convolution, each feature map is summed point-wise with the corresponding convolved ``modification probability map'' (see Figure \ref{fig:SCA-spatial}). Note that the activation functions of the first convolutions in SCA-Ye-Net, i.e. the truncation activation function ({\it truncated linear unit (TLU)} in the article), are replaced by a ReLU. This makes it possible to propagate (forward pass) ``virtually'' throughout the network, an information related to the image, and another related to the modification probability map.

Note that this procedure to transform a Not-SCA-CNN into an SCA-CNN is inspired by the propagation of the modification probability map proposed in \cite{Denemark2016_SCA_Sigma} and \cite{Denemark2016_SCA_Delta}. These two papers come as an improvement on the previous maxSRM Rich Models \cite{Denemark2014_maxSRM}. In maxSRM, instead of accumulating the number of occurrences in the co-occurrence matrix, an accumulation of the maximum of a local probability was used. In \cite{Denemark2016_SCA_Sigma} and \cite{Denemark2016_SCA_Delta}, the idea was to transform the modification probability map in a similar way to the filtering of the image, and then to update the co-occurrence matrix using the transformed version of the modification probability map, instead of the original modification probability map. The imitation of this principle was initially integrated into Ye-Net for CNN steganalysis, and this concept is easily transposable to most of the modern CNNs.

\subsection{The JPEG steganalysis}
\label{ssec:jpeg_nsca}

The best JPEG CNN at the end of 2018 was SRNet \cite{Boroumand2018_SRNet}. Note that this network, at this period, is the only one that has been proposed with a Side Channel Aware (SCA) version. 

It is interesting to list and rapidly discuss the previous CNNs used for JPEG steganalysis. The first network, published in February 2017, was the Zeng {\it et al.} network and was evaluated with a million images, and does a limited evaluation of stego-mismatch \cite{Zeng2017} \cite{Zeng2017_Millions}. Then in June 2017 at IH\&MMSec'2017, two networks have been proposed: PNet \cite{Chen2017_PNet}, and Xu-Net-Jpeg \cite{Xu2017}. Finally, SRNet \cite{Boroumand2018_SRNet} was added online in Septembre 2018.

In Zeng {et al.}s' network \cite{Zeng2017} \cite{Zeng2017_Millions}, the pre-processing block takes as input a de-quantized (real value) image, then convolved it with 25 DCT basis, and then quantized and truncated the 25 filtered images. This pre-processing block, uses handcrafted filter kernels (DCT basis), the kernels' values are fixed, and these filters are inspired by DCTR Rich Models \cite{Holub2015_DCTR}. There are three different quantizations, so, the pre-processing block gives $3\times25$ residual images. The CNN is then made of 3  sub-networks which are each producing a feature vector of 512 dimension. The  sub-networks are inspired by Xu-Net \cite{Xu2016a}. The three feature vectors, outputed by the three  sub-networks, are then given to a fully connected structure, and the final network ends with a softmax layer.

Similarly to what has been done for spatial steganalysis, this network is using a pre-processing block inspired by Rich Models \cite{Holub2015_DCTR}. Note that the most efficient Rich Models today is the Gabor Filter Rich Models \cite{Xia2017_GFR}. Also, note that this network takes advantage of the notion of an ensemble of features, which comes from the three different sub-networks. The network of Zeng {et al.} is less efficient than Xu-Net-Jpeg \cite{Xu2017}, but gives an interesting first approach guided by Rich Models. 

The PNet main idea (and also VNet which is less efficient but takes less memory) \cite{Chen2017_PNet} is to imitate Phase-Aware Rich Models, such as DCTR \cite{Holub2015_DCTR}, PHARM \cite{Holub2015_PHARM}, or GFR \cite{Xia2017_GFR}, and therefore to have a decomposition of an input image into 64 features maps which represents the 64 phases of the Jpeg images. The pre-processing block takes as input a de-quantized (real value) image, convolves it with four filters, the ``SQUARE5$\times$5'' from the Spatial Rich Models \cite{Fridrich2012_Rich}, a ``point'' high-pass filter (referenced as ``catalyst kernel'') which complements the ``SQUARE5$\times$5'', and two directional Gabor Filters (angles 0 and $\pi/2$). 

Just after the second block of convolution, a ``PhaseSplit Module'' splits the residual image into 64 feature maps (one map = one phase), similarly to what was done in Rich Models. Some interesting methods have been used such as (1) the succession of the fixed convolutions of the pre-processing block, and a second convolution with learnable values, (2) a clever update of BN parameters, (3) the use of the ``Filter Group Option'' which virtually builds sub-networks, (4) bagging on 5-cross-validation, (5) taking the 5 last evaluations in order to give the mean error for a network, (6) shuffling the database at the beginning of each epoch, to have better BN behavior, and to help generalization, and (7) eventually using an Ensemble. With such know-how, PNet beat the classical two-step machine learning approaches in a Not-SCA, and also in a SCA version (Ensemble Classifier + GFR).

The Xu-Net-Jpeg \cite{Xu2017} is even more attractive since the approach was slightly better than PNet, and does not require a strong domain inspiration like in PNet. The Xu-Net-Jpeg is strongly inspired by ResNet \cite{He2016_ResNet}, a well-established network from the machine learning community. ResNet allows the use of deeper networks thanks to the use of shortcuts. In Xu-Net-Jpeg, the pre-processing block takes as input a dequantized (real value) image, then convolves the image with 16 DCT basis (in the same spirit as Zeng {et al.} network \cite{Zeng2017} \cite{Zeng2017_Millions}), and then applies an absolute value, a truncation, and a set of convolutions, BN, ReLU until it obtains a feature vector of 384 dimension, which is given to a fully connected block. We can note that the max pooling or average pooling are replaced by convolutions. This network is really simple and was in 2017, the state-of-the-art method. In a way, these kind of results shows us that the networks proposed by machine learning community are very competitive and there is not so much domain-knowledge to integrate to the topology of a network in order to obtain a very efficient network.

In 2018 the state-of-the-art CNN for JPEG steganalysis (which can also be used for spatial steganalysis) was SRNet \cite{Boroumand2018_SRNet}. This network was previously presented in Section \ref{ssec:nsca}. Note that for the side channel aware version of SRNet, the {\it embedding change probability} per DCTs coefficient is, first, mapped back in the spatial domain using absolute values for the DCT basis. This {\it side-channel map} then enters the network and is convolved with each kernel (this first convolution acts as a pre-processing block). Note that the convolutions in this first block for this {\it side-channel} map are such that the filter kernels are modified to their absolute values. After passing the convolution, the feature maps are summed with the square root of the values from the convolved {\it side-channel} map. Note that this idea is similar to what was exposed in SCA Ye-Net version (SCA-TLU-CNN) \cite{Ye2017} about the integration of a Side-Channel map, and to the recent proposition for Side-Channel Aware steganalysis in JPEG with Rich Models \cite{Denemark2016_SCA_Delta}, where the construction of the {\it side-channel} map, and especially the quantity $\delta_{uSA}^{1/2}$\footnote{uSA stands for Upper bounded Sum of Absolute values.} was defined.

Note that a similar solution with more convolutions, applied to the {\it side-channel map}, have been proposed in IH\&MMSec'2019 \cite{Huang2019_LMC}.

\subsection{Discussion about the Mismatch phenomenon scenario}
\label{ssec:mismatch}

Mismatch (cover-source mismatch or stego-mismatch) is a phenomenon present in machine learning, and this issue sees classification performances decrease because of the inconsistency between the distribution of the learning database and the distribution of the test database. The problem is not due to an inability to generalize in machine learning algorithms, but due to the lack of similar examples occurring in the training and test database. The problem of mismatch is an issue that goes well beyond the scope of steganalysis.

In steganalysis the phenomenon can be caused by many factors. The cover-source mismatch can be be caused by the use of different photo-sensors, by different digital processing, by different camera settings (focal length, ISO, lens, etc), by different image sizes, by different image resolutions, etc \cite{Gibouloto2018_CSM}, \cite{Borghys2018_CSM}. The stego-mismatch can be caused by different amounts of embedded bits, or by different embedding algorithms.

Even if not yet fully explored and understood, the mismatch (cover-source mismatch (CSM) or stego mismatch) is a major area for examination in the coming years for the discipline. The results of the Alaska challenge~\cite{Cogranne2019_Alaska}~\footnote{Alaska: A challenge of steganalysis into the wilderness of the real world. \url{https://alaska.utt.fr/}.} published at the ACM conference IH\&MMSec'2019 will continue these considerations.

In 2018, CSM had been established for 10 years \cite {Cancelli2008}. There are two major current school of thought, as well as a third more exotic one:
\begin{itemize}
\item The first school of thought is the so-called {\bf holistic} approach (that is to say, global, macroscopic, or systemic), and consists of learning all distributions \cite{Lubenko2012_Perceptron}, \cite{Lubenko2012_LargeData}. The use of a single CNN with millions of images \cite{Zeng2017_Millions} is in the logical continuation of this current school of thought. Note that this scenario does not consider that the test set can be used during learning. This scenario can be assimilated to an {\it online scenario} where the last player (from a game theory point of view) is the steganographer because in an online scenario the steganographer can change her strategy while the steganalyzer cannot.
\item The second school of thought is {\bf atomistic} (= partitioned, microscopic, analytical, of divide-and-conquer type, or individualized) and consists of partitioning the distribution \cite{Pasquet2014_Islet}, that is to say to create a partition and to associate a classifier for each cell of the partition. Note that an example of an atomistic approach for stego-mismatch management, using a CNN multi-classifier, is presented in \cite{Butora2019_StegoMultiClass} (a class is associated with each embedding algorithm - there is thus a latent partition). Note that this idea \cite{Butora2019_StegoMultiClass}, among others, has been used by the winners of the Alaska challenge \cite{Yousfi2019_BreakingALASKA}. Note that again, this scenario does not consider that the test set can be used during learning. This scenario can also be assimilated to an {\it online scenario} where the last player (from a game theory point of view) is the steganographer because in an online scenario the steganographer can change her strategy while the steganalyst cannot.
\item Finally, the third exotic school of thought considers that there is a test database (with much more than one image), and that the database is available, and usable (without labels) during learning. This scenario can be assimilated to an {\it offline scenario} where the last player (from a game theory point of view) is the steganalyser, because in this offline scenario the steganalyser is playing a more forensic role. In this situation, there are approaches of type domain adaptation, or a transfer of features GTCA \cite{Li2013_GTCA}, IMFA \cite{Kong2016_IMFA}, CFT\cite{Feng2017_CFT}, where the idea is to define an invariant latent space. Another approach is ATS \cite{Lerch-Hostalot2016_ATS} which performs an unsupervised classification using only the test database and requires the embedding algorithm in order to re-embed a payload in the images from the test database.
\end{itemize}

These three schools of thought can help derive approaches by CNN that integrate the ideas presented here. That said, the ultimate solution may be to detect the phenomenon of mismatch and raise the alarm or prohibit the decision \cite{Koak2017_SafePredict}. In short, to integrate a more intelligent mechanism than just holistic or atomistic.

%%%%%%%%%%%%%%%%%%%%%%%%%%%%%%%%%%%%%%%%%%%%%%%%%%%%%%%%%%%%%%%%%%%%%%%%%%%%%%%%%%%%%%%%%%%%%%%%%%%%%
% SECTION
%%%%%%%%%%%%%%%%%%%%%%%%%%%%%%%%%%%%%%%%%%%%%%%%%%%%%%%%%%%%%%%%%%%%%%%%%%%%%%%%%%%%%%%%%%%%%%%%%%%%%
\section{Steganography by deep-learning}
\label{sec:GAN}

In Simmons' founding article \cite{Simmons1983}, steganography and steganalysis are defined as a {\it 3-player game}. The steganographers, usually named Alice and Bob, want to exchange a message without being suspected by a third party. They must use a harmless medium, such as an image, and hide the message in this medium. The steganalyst, usually called Eve, observes the exchanges between Alice and Bob. Eve must check whether these images are natural, that is to say, cover images, or whether they hide a message, i.e. stego images.

This notion of {\it game} between Alice, Bob and Eve corresponds to that found in game theory. Each player tries to find a strategy that maximizes their chances of winning. For this, we express the problem as a min-max problem that we seek to optimize. The solution to the optimum, if it exists, is called the solution at the Nash equilibrium. When all the players are using a strategy at the Nash equilibrium, any change of strategy from a player, leads to a counter attack from the other players allowing them to increase their gains.

In 2012, Sch\"{o}ttle and B\"{o}hme \cite{SchottleIH2012_Game}, \cite{SchottleTIFS2016_Game} have modeled with a simplifying hypotheses a problem of steganography and steganalysis and proposed a formal solution. Sch\"{o}ttle and B\"{o}hme have named this approach the {\it optimum adaptive steganography} or {\it strategic adaptive steganography} in opposition to the so-called {\it naive adaptive steganography} that corresponds to what is currently used in algorithms like HUGO (2010) \cite{Pevny2010-HUGO}, WOW (2012) \cite{Holub2012_WOW}, S-UNIWARD / J-UNIWARD / SI-UNIWARD (2013) \cite{Holub2014_UNIWARD}, HILL (2014) \cite{Li2014_HILL}, MiPOD (2016) \cite{Sedighi2016_MIPOD}, Synch-Hill (2015) \cite{Denemark2015_Synchronizing}, UED (2012) \cite{Guo2012_UED}, IUERD (2016) \cite{Pan2016_IUERD}, IUERD{\it -UpDist-Dejoin2} (2018) \cite{Li2018_IUERD-UpDist-Dejoin2}, etc.

That said, the mathematical formalization of the steganography / steganalysis problem by game theory is difficult and often far from practical in reality. Another way to determine a Nash equilibrium is to ``simulate'' the game. From a practical point of view, Alice plays the entire game alone, meaning that she does not interact with Bob or Eve to build her embedding algorithm. The idea is that she uses 3 algorithms (2 algorithms in the simplified version) that we name {\it agents}. Each of these agents will play the role of Alice, Bob \footnote{Bob is deleted in the simplified version.} and Eve, and each agent runs at Alice's home. Let us note these three algorithms running at Alice's home: {\it Agent-Alice}, {\it Agent-Bob}, and {\it Agent-Eve}. With these notations, we thus make a distinction with the Human users: Alice (sender), Bob (receiver), and Eve (warden), and it allows us to highlight the fact that the three agents are executed from Alice's side. So, Agent-Alice's role is to embed a message into an image so that the resulting stego image is undetectable by Agent-Eve, and such that Agent-Bob can extract the message. 

Alice can launch the game, that is to say the simulation, and the agents are ``fighting''~\footnote{The reader should be aware that from a game theory point of view there are only two teams that are competing (Agent-Alice plus Agent-Bob from one side, and Agent-Eve from the other side) in a zero-sum game.}. Once the agents have reached a Nash's equilibrium, Alice stops the simulation and can now keep Agent-Alice, which is her {\it strategic adaptive embedding} algorithm, and can send Agent-Bob i.e the extraction algorithm (or any equivalent information) to Bob \footnote{Note that the exchange of any secret information between Alice and Bob, prior to the use of Agent-Alice and Agent-Bob, requires the use of another steganographic channel. Also note that this initial sending from Alice to Bob before been able to use Agent-Alice and Agent-Bob is equivalent to the classical stego-key exchange problem.}. The secret communication between Alice and Bob is now possible through the use of the Agent-Alice algorithm for embedding and Agent-Bob algorithm for the extraction.

\begin{figure*}[ht]
\centering
  \FIG{\includegraphics[width=7cm,keepaspectratio=False]{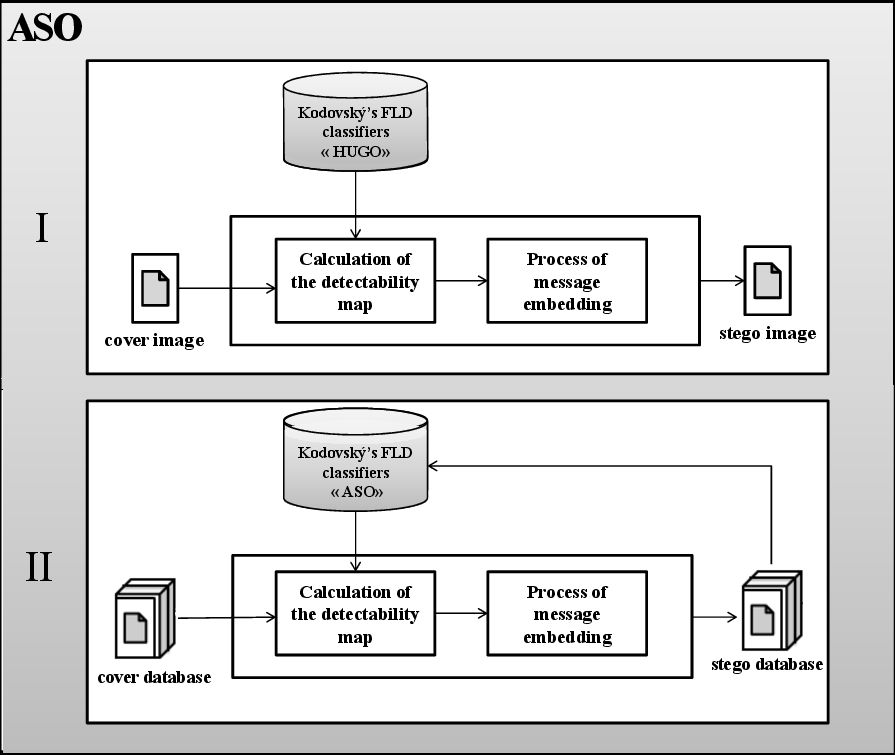}}
      {\caption{General scheme of ASO \cite{Kouider2013} \cite{Kouider2012}.}}
  \label{fig:aso}
\end{figure*}

The first precursor approaches aimed at simulating a {\it strategic adaptive equilibrium}, and therefore proposing {\it strategic embedding} algorithms date from 2011 and 2012. The two approaches are MOD \cite{Filler2011_MOD} and ASO \cite{Kouider2013} \cite{Kouider2012}; See Figure \ref{fig:aso}. Whether for MOD or ASO, the game is made by pitting Agent-Alice and Agent-Eve against each other. In this game, Agent-Bob is not used since Agent-Alice is simply generating a cost map, which is then used for coding and embedding the message thanks to an STC \cite{Filler2011STC}. Alice can generate a cost map for a source image with the Agent-Alice, and then she can easily use the STC \cite{Filler2011STC} algorithm to embed her message and obtain the stego image. From his side, Bob only has to use the STC \cite{Filler2011STC} algorithm to retrieve the message from the stego image. 

In both MOD or ASO, the ``simulation'' is such that the two following actions are iterated until a stop criterion is reached:
\begin{enumerate}[label=\roman*)]
\item Agent-Alice updates its embedding cost map by asking an Oracle (the Agent-Eve) how best to update each embedding cost, to be even less detectable. 
\begin{description}
\item [In MOD (2011) \cite{Filler2011_MOD},] Agent-Eve is an SVM. Agent-Alice updates their embedding costs by reducing the SVM margin separating the covers and the stegos. 
\item [In ASO (2012) \cite{Kouider2013},] Agent-Eve is an Ensemble Classifier \cite{Kodovsky2012-EnsembleClassifiers} and is named an Oracle. Agent-Alice updates their embedding costs by transforming a stego in a cover. 
\end{description}
In both cases, the idea is to find a displacement in the latent space (feature space) co-linear to the orthogonal axis to the hyperplane separating the cover and stego class. Note that in the current terminology, introduced by Ian Goodfellow in 2014 \cite{Goodfellow2014_GAN}, Agent-Alice runs an adversarial attack, and the Oracle (Agent-Eve), named a discriminator (or the classifier to be deceived), must learn to counter this attack.
\item The Oracle (Agent-Eve) updates its classifier. Reformulated with the terminology from machine learning, this equates to the discriminant update by re-learning it, in order to steganalysis once more the stego images generated by Agent-Alice.
\end{enumerate}

In 2014, Goodfellow {\it et al.} \cite{Goodfellow2014_GAN} used neural networks to ``simulate'' a game with an {\it image generator network} and a {\it discriminating network} whose role was to decide whether an image was real or synthesized. The authors have named this Generative Adversarial Networks (GAN approach). The terminology used in this paper was subsequently widely adopted. Moreover, the use of neuron networks makes the expression of the min-max problem easy. The optimization is then carried out via the back-propagation optimization process. Moreover, thanks to deep-learning libraries it is now easy to build a GAN type system. As we have already mentioned before, the concept of game simulation, existed in steganography / steganalysis with MOD \cite{Filler2011_MOD} and ASO \cite {Kouider2013}, but the implementation and the optimization becomes easier with neural networks.

From 2017, after a period of 5 years of stagnation, the concept of the simulated game is once again studied in the field of steganography / steganalysis, thanks to the emergence of deep learning and GAN approaches. At the end of 2018, we can define four groups or four families~\footnote{\samepage ``Deep Learning in Steganography and Steganalysis since 2015'', tutorial given at the ``Image Signal \& Security Mini-Workshop'', the 30th of October 2018, IRISA / Inria Rennes, France, DOI: 10.13140/RG.2.2.25683.22567, {\it http://www.lirmm.fr/~chaumont/publications}. See the slides \href{http://www.lirmm.fr/~chaumont/publications/Deep_Learning_in_Steganography_and_Steganalysis_since_2015_Tutorial_Meeting-France-CHAUMONT_30_10_2018.pdf}{here}, and the video of the talk \href{https://videos-rennes.inria.fr/video/H1YrIaFTQ}{here}.} of approaches; some of which will probably merge:
\begin{itemize}
\item The family by synthesis,
\item The family by generation of the modifications probability map,
\item The family by adversarial-embedding {\it iterated} (approaches misleading a discriminant),
\item The family by 3-player game,
\end{itemize}

\subsection{The family by synthesis}

The first approaches based on {\it image synthesis} via a GAN \cite{Goodfellow2014_GAN} generator proposed the generation of cover images and then use them to make insertion by modification. These early propositions were approaches {\it by modification}. The argument put forward for such approaches is that the generated database would be safer. A reference often cited is that of SGAN \cite{Volkhonskiy2017_SGAN} found on ArXiv, which was rejected at ICLR'2017 and was subsequently never published. This unpublished paper has a lot of errors and lack of proof. We should rather prefer the reference of SSGAN \cite{Shi2017_SSGAN} that was published in September 2017, and that proposes the same thing: generate images and then hide messages in them. However, this protocol seems to complicate the matter. It is more logical that Alice herself chooses natural images that are safe for embedding, i.e. images that are innocuous, never broadcasted before, adapted to the context, with lots of noise or textures \cite{Sedighi2016_TossBOSS}, not well classified by a classifier \cite{Kouider2012} or with a small deflection coefficient \cite{Sedighi2016_MIPOD}, rather than generating images, and then using them to hide a message. 

\begin{figure*}[ht]
\centering
  \FIG{\includegraphics[width=\columnwidth,keepaspectratio=False]{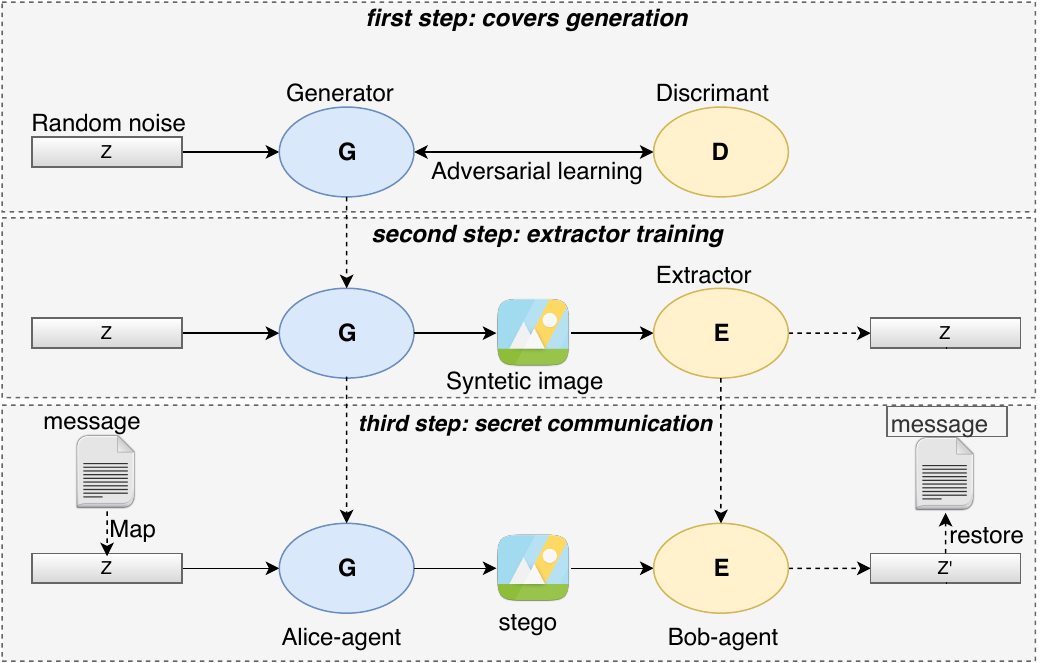}}
      {\caption{Hu {\it et al.} \cite{Hu2018_DCGAN_SWE} approach by synthesis without modification.}}
  \label{fig:swe}
\end{figure*}

A much more interesting approach using {\it synthesis} is to directly generate images that will be considered stego. To my knowledge, the first approach exploiting the GAN mechanism for image synthesis using the principle of steganography {\it without modifications} \cite{Fridrich2009_Book} is proposed in the article of Hu {\it et al.} \cite{Hu2018_DCGAN_SWE} and published in July 2018; See Figure \ref{fig:swe}.

%\begin{figure*}[!htb]
%\centering
  %\FIG{\includegraphics[width=\columnwidth]{Yedrouj-Net.png}}
      %{\caption{Comparison of Yedroudj-Net, Xu-Net, and Ye-Net architectures.}}
%\label{fig:nets}
%\end{figure*}

The first step consists of deriving a network able to synthesize images. In this paper, the DCGAN generator \cite{Radford2015_DCGAN} is used to synthesize images with a preliminary learning thanks to GAN methodology. When fed with a vector of a fixed-size uniformly distributed in $[-1, 1]$ the generator synthesizes an image. The second step consists of learning to another network to extract a vector from a synthesized image; the extracted vector must correspond to the vector given at the input of the generator which synthesizes the image. Finally, the last step consists of sending Bob the extraction network. Now, Alice can map a message to a fixed-size uniformly distributed vector, and then synthesize an image with the given vector, and send it to Bob. Bob can extract the vector and retrieve the corresponding message.

The approaches with {\it no modifications} have been around for many years, and it is known that one of the problems is that the number of bits that can be communicated is lower compared to the approaches with modifications. That said, the gap between the approaches by {\it modifications} versus {\it no-modifications} is beginning to narrow.

Here is a rapid analysis of the efficiency of the method. In the paper of Hu {\it et al.} \cite{Hu2018_DCGAN_SWE}, the capacity is around 0.018 bits per pixel (bpp) with images $64\times64$ pixels~\footnote{The vector dimension is 100. This vector is used to synthesize images of a size $64\times64\times3$. There are $100 \times 3$ bits (see the mapping) per image, i.e. about 0.02 bits per pixel (bpp). The Bit Error Rate is BER $= 1 - 0.94 = 6$\%. It is, therefore, necessary to add an Error Correcting Code (ECC) so that the approach is without errors. With the use of a Hamming code $[15,11,3]$ that corrects at best 6\% of errors, the payload size is therefore around 0.018 bpp.}. In the experiment carried out, the synthesized images are either faces or photographs of food. An algorithm like HILL\cite{Li2014_HILL} (one of the most powerful algorithms on the BOSS database \cite{Sedighi2016_TossBOSS}) is detected by SRNet \cite{Boroumand2018_SRNet} (one of the most successful steganalysis approaches towards the end of 2018) with a probability of error of Pe = 31.3\% (note that a Pe of 50\% is equivalent to a random detector) on a $256\times256$ pixels BOSS database, for a payload size of 0.1 bpp. Due to the square root law, the Pe would be higher for the $64\times64$ pixels BOSS database.

Therefore, there is around 0.02 bpp for the unmodified synthetic approach of Hu {\it et al.} \cite{Hu2018_DCGAN_SWE} whose security has not yet been evaluated enough, against something around 0.1 bpp for HILL, with less than one chance in three to be detected with a {\it clairvoyant} steganalysis i.e. a laboratory steganalysis (unrealistic and much more efficient than a ``real-world'' / ``into the wild'' steganalysis \cite{Ker2013_RealWorld} \cite{Cogranne2019_Alaska}). Therefore, there is still a margin in terms of the number of bits transmitted between the {\it no-modification} synthesis-based approaches, such as that of Hu {\it et al.} approach \cite{Hu2018_DCGAN_SWE}, and {\it modification} approaches such as S-UNIWARD \cite{Holub2014_UNIWARD}, HILL \cite{Li2014_HILL}, MiPod \cite{Sedighi2016_MIPOD} or even Synch-Hill \cite{Denemark2015_Synchronizing}, but this margin has been reduced~\footnote {The other families of steganography by deep learning, which are {\it modification} based, will probably help to maintain this performance gap for a few years more.}. Also, note that there are still some issues to be addressed to ensure that approaches such as the one proposed by Hu {\it et al.} are entirely safe. In particular, it must be ensured that the detection of synthetic images \cite{Quan2018_DiscinctionSynthesis} does not compromise the communication channel in the long term. It must also be ensured that the absence of a secret key does not jeopardize the approach. Indeed, if one considers that the generator is public, is it possible to use this information to deduce that a synthesis approach without modification has been used?

\subsection{The family by generation of the modifications probability map}

\begin{figure*}[ht]
\centering
  \FIG{\includegraphics[width=\columnwidth,keepaspectratio=False]{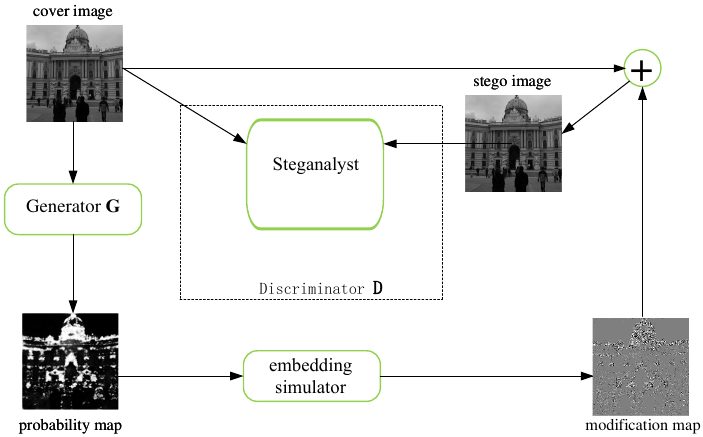}}
      {\caption{ASDL approach; generation of the modifications probability map}}
  \label{fig:ASDL}
\end{figure*}

The family by generation of the modification probability map is summarized in late 2018 in two papers: ASDL-GAN \cite{Tang2017_ASDL-GAN}, and UT-6HPF-GAN \cite{Yang2018_UT-6HPF-GAN}; See Figure \ref{fig:ASDL}. In this approach, there is a generator network and a discriminant network. From a cover, the generator network generates a map which is named the modification probability map. This modification probability map is then passed to an equivalent of the random draw function used in the STC \cite{Filler2011STC} simulator. We then obtain a map whose values ​​belong to \{-1, 0, +1\}. This map is called the modification map and corresponds to the so-called stego-noise. The discriminant network takes as input a cover or an image resulting from the summation (point-to-point sum) of the cover and the stego-noise generated by the generator. The discriminant's objective is to distinguish between the cover and the ``{\it cover $+$ stego-noise}'' image. The generator's objective is to generate a modification map which makes it possible to mislead the discriminant the most. Of course, the generator is forced to generate a non-zero probability map by adding in the loss term, a term constraining the size of the payload in addition to the term misleading the discriminant.

In practice, taking the latest approach UT-6HPF-GAN \cite{Yang2018_UT-6HPF-GAN}, the generator is a U-Net type network, the draw function is obtained by a differentiable function {\it double Tanh}, and the discriminant is the Xu-Net \cite{Xu2016a} enriched with 6 high-pass filters for the pre-processing in the same spirit as Ye-Net \cite{Ye2017} or Yedroudj-Net \cite{Yedroudj2018_Net}.

The system learns on a first database, and then security comparisons are made on the $256\times256$ pixels BOSS \cite{Bas2011-BOSS}, LIRMMBase \cite{Pibre2016}, and BOWS2 \cite{BOWS2008} databases. The steganalysis is done with the 
Ensemble Classifier (EC) \cite{Kodovsky2012-EnsembleClassifiers} fed by SRM \cite{Fridrich2012_Rich}, with EC plus the MaxSRM \cite{Denemark2014_maxSRM}, and with Xu-Net \cite{Xu2016a}. Note that using Xu-Net is not a good choice since it is less efficient than EC+SRM or EC+MaxSRM, and also because it is the discriminant in the UT-6HPF-GAN (there is a risk of falling into an ``incompleteness'' issue; see papers \cite{Kodovsky2008} \cite{Kodovsky2011_dangers}). So, only looking at the results with EC+SRM, on the BOSS database, with real embedding using STC \cite{Filler2011STC}, the performances are equivalent to those of HILL \cite{Li2014_HILL}, which is one of the most efficient embedding algorithms on BOSS \cite{Sedighi2016_TossBOSS}. It is therefore a very promising family. 

Additionally, the generator does not seem to be impacted when used on a database that is different from the learning database. Nevertheless, curriculum learning has to be used when the target payload is changed, which seems to indicate a kind of sensitivity to the mismatch. Further reflexions have also to be achieved related to the generator's loss, and the mixing of both a security-related term and a payload-size term. Usually, one of the two criteria is fixed, so that we have to be in a payload-limited sender scenario or a security-limited sender scenario. Note that a version for JPEG has been proposed in IH\&MMSec'2019, JS-GAN \cite{Yang2019_JS-GAN}.

\subsection{The family by adversarial-embedding {\it iterated} (approaches misleading a discriminant)}

The family by adversarial-embedding \underline{iterated} re-uses the concept of {\it game simulation} which was presented in the beginning of Section \ref{sec:GAN} with a simplification of the problem since there are only two-players: Agent-Alice and Agent-Eve. Historically MOD \cite{Filler2011_MOD} and ASO \cite{Kouider2013} were the first algorithms of this type. 

Recently some papers have used the adversarial concept\footnote{An adversarial attack does not necessarily require us to use a deep learning classifier.} by generating a deceiving example (see \cite{Zhang2018_Adverserial}), but these approaches are not adversarial-embedding \underline{iterated}. Nor are they dynamic, they contain no game simulation, they are not trying to reach a Nash equilibrium, there is no learning alternation between the embedder and the steganalysis.

A paper whose spirit is more in tune with a simulation of a game, which takes the principle of ASO \cite{Kouider2013}, and whose objective is to update the cost map is the algorithm ADV-EMB \cite {Tang2019_ADV-EMB} (previously named AMA on {\it ArXiv} \href{https://arxiv.org/abs/1803.09043}{arXiv:1803.09043}). In this article, the authors propose to make an adversarial-embedding {\it iterated}, by letting Agent-Alice access the gradient of the loss of Agent-Eve (similarly to ASO, where Agent-Alice has access to its Oracle (the Agent-Eve)). In ADV-EMB, Agent-Alice uses the gradient, of the direction to the class frontier (between classes cover and stego), to modify the cost map, and in ASO, Agent-Alice directly uses the direction of the class frontier to modify the cost map.

In ADV-EMB \cite{Tang2019_ADV-EMB}, the cost map is initialized with the cost of in S-UNIWARD (for ASO it was the cost of HUGO \cite{Pevny2010-HUGO}). During the iterations, the cost map is updated, but there is only a $\beta$ percentage of values that are updated \footnote{In STC, before coding the message, the pixels position of the image are shuffled thanks to the use of a pseudo-random shuffler, seeded by the secret stego-key. Note that this stego-key is shared between Alice and Bob. After the shuffling step, ADV-EMB selects the last $\beta$ percent pixels of the {\it shuffled} image, and modifies their associated cost and only those ones.}. When the ADV-EMB iterations are stopped, the cost map is composed of a $\beta-1$ percent of positions having a cost defined by S-UNIWARD, and $\beta$ percent of positions having a cost coming from a change in the initial cost given by S-UNIWARD.

Note that updating a cost causes a cost asymmetry since the cost of a +1 change is no longer equal to the cost of a -1 change, as in ASO. Besides, the update of the two costs of a pixel is rather rough since it is a simple division by 2 for a direction (+1 or -1) and multiplication by 2 for the other direction. The sign of the gradient of loss, calculated by choosing the cover label, for a given pixel, makes it possible to determine for each of the two directions (+ 1 / -1) if we should reduce or increase the cost. The idea is as in ASO, to deceive the discriminant since when we decide to reduce the value of a cost, it is to favor the direction of modification associated with this cost, and thus we promote getting closer to the cover class.

With such a scheme, security is improved. The fact that it is preferable to have a small number of modifications to the initial cost map probably makes it possible to preserve the initial embedding approach, and therefore not to introduce too many traces that could be detected by another steganalyzer \cite{Kodovsky2011_dangers}. That said, the update to the costs should probably be refined to better take into account the value of the gradient. For the moment, the selection of the $\beta$ percent of pixels that will be modified is suboptimal, and this selection should eventually be done by looking at the initial cost of the whole pixel. Finally, as it is the case for ASO, if the discriminant is not powerful enough to carry out a steganalysis, then it can be totally counterproductive for the Agent-Alice. Therefore, there are many open questions regarding the convergence criterion, the stopping criterion, the number of iterations in the alternation between Agent-Alice and Agent-Eve, and the definition of a metric for measuring the relevance of Agent-Eve, etc. Note that an adversarial embedding {\it iterated} with Agent-Alice countering multiple versions of Agent-Eve has been proposed in IH\&MMSec'2019 \cite{Bernard2019_AdverserialEmbedding}.

\subsection{The family by 3-player game}

\begin{figure*}[tb]
  \FIG{\includegraphics[width=\columnwidth]{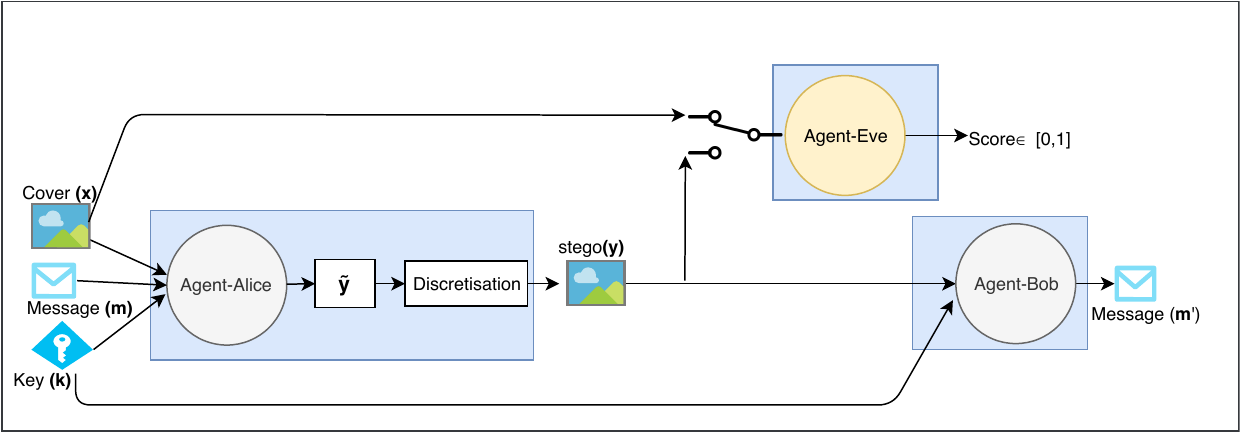}}
		  {\caption{The overall architecture of the {\it 3-players game}.}}
  \label{fig:3-players}
\end{figure*}

The 3-player game concept is an extension of the previous family (see the family ``adversarial-embedding {\it iterated}''), but this time with three agents and all are neural networks. Here,  the three agents: Agent-Alice, Agent-Bob, and Agent-Eve are present (see Section \ref{sec:GAN} for an overview of the game). Note that Agent-Alice and Agent-Bob are ``linked'' since Agent-Bob is only there to add a constraint on the solution obtained by Agent-Alice. Thus, the primary ``game'' is an antagonistic (or adversarial) game between Agent-Alice and Agent-Eve, while the ``game'' between Agent-Alice and Agent-Bob is rather cooperative, since these two agents share the common purpose of communicating (Agent-Alice and Agent-Bob both want Agent-Bob to be able to extract the message without errors). Figure \ref{fig:3-players}, from \cite{Yedroudj2019_3players} summarizes the principle of the 3-player game. Agent-Alice takes a cover image, a message and a stego-key, and after a discretisation step generates a stego image. This stego image is used by Agent-Bob to retrieve a message. On the other side, Agent-Eve has to decide whether an image is cover or stego; this agent outputs a score.

Historically, after MOD and ASO, which only included two players, we can see the premise of the idea of three players appear in 2016 with the paper of Abadi and Andersen \cite{Abadi2016_AdversarialCrypto}. In this paper, Abadi and Andersen \cite{Abadi2016_AdversarialCrypto} from Google Brain, proposed a cryptographic toy-example for an encryption based on the use of three neural networks. The use of neural networks makes it easy to obtain a {\it strategic equilibrium} since the problem is expressed as a min-max problem and its optimization can be carried out by the back-propagation process. Naturally, this 3-player game concept can be transposed to steganography with the use of deep learning.

In December 2017 (GSIVAT; \cite{HayesNIPS2017_3players}), and in September 2018 (HiDDeN; \cite {ZhuECCV2018_3players}), two different teams from the machine learning community proposed, in NIPS'2017, then in ECCV'2018, to achieve {\it strategic embedding} thanks to 3 CNNs, iteratively updated, who play the role of Agent-Alice, Agent-Bob, and agent Agent-Eve. These two articles do not rigorously define the concept of the 3-player game, and there are erroneous assertions, mainly because the security and its evaluation are not correctly handled. If we place ourself in the standard framework to evaluate the empirical security of an embedding algorithm, that is to say with a clairvoyant Eve, the two approaches are very detectable. The most significant issues with these two papers are first, neither of the two approaches uses a stego-key; which is the equivalent to always using the same key, and it leads to very detectable schemes \cite{Pibre2016}, second, there is no discretization of pixel values issued from Agent-Alice, third, the computational complexity, due to the use of fully connected blocks, leads to un-practical approaches, and fourthly, the security evaluation is not carried out with a state-of-the-art steganalyzer. 

At the beginning of 2019, Yedroudj {et al.} \cite {Yedroudj2019_3players} redefined the 3-player concept, by integrating the possibility of using a stego-key, treating the problem of discretization, going through convolution modules to have a scalable solution, and using a suitable steganalyzer. The proposition is not comparable to classical adaptive embedding approaches, but there is a real potential to such an approach. The Bit Error Rate is sufficiently small to be nullified, the embedding is done in the texture parts, and security could be improved in the future. As an example, the probability of error with a steganalysis by Yedroudj-Net\cite{Yedroudj2018_Net}, under equal errors prior, for a real payload size 0,3 bpp\footnote{A Hamming error correcting code ensures a null BER theoretically for most of the images, and thus a rate of 0.3 bpp for these images.} for images of from BOWS2 database is 10.8\%. This can, for example, be compared to the steganalysis of WOW\cite{Holub2012_WOW} using the same conditions, which give a probability error of 22.4\%. There is still a security gap, but this approach paves the way to much research. There are still open questions on the link between Agent-Alice and Agent-Bob, on the use of GANs, and on the definition of losses and the tuning of the compromises between the different constraints.

%%This architecture, with a real payload size of 0,3 bpp, can retrieve perfectly an embedded message (with the help of error correcting code), for security level Pe = 10,8%

%\pagebreak

\section*{Conclusion}

In this chapter, we have practically completed a full presentation of the subject on deep learning in steganography and steganalysis, since its appearance in 2015. As a reviewer of many papers related to this subject during the period 2015 - 2018, I think and I hope this chapter will help the community to understand what has been done and what are the next directions to explore.

In this chapter, we recalled the main elements of a CNN. We discussed the memory and time complexity, and practical problems for efficiency. We explored the link between some past approaches sharing similarities with what is currently carried out in a CNN. We presented the various main networks until the beginning of 2019, and multiple scenarios, finally we touched on the recent approaches for steganography with deep learning.

As mentioned in this chapter, many things have not been solved yet, and the major issue is to be able to experiment with more realistic hypotheses to be more ``into the wild''. The ``holy grail'' is cover-source mismatch and stego-mismatch, but in a way, the mismatch is a problem shared by the whole machine learning community. CNNs are now very present in the steganalysis community, and the next question is probably: how to go a step further and produce clever networks?

\begin{backmatter}

%%%% To be used for Chapter-wise bibliography with the help of "chapterbib" package.
%%%% Also use the natbib package option "sectionbib" with your template.
%%%% The "chapterbib" package and its documentations are available in:
%%%% 	.
%\pagebreak
\section*{REFERENCE}
\addcontentsline{toc}{section}{Reference}
\nopagebreak
%\nopagebreak
%\bibliographystyle{alpha}
%\bibliographystyle{./chapitre/IEEEtran}
%\bibliographystyle{./chapitre/acm}
%\bibliographystyle{plainnat-fr}
%\bibliographystyle{plainnat-en}
%\bibliographystyle{Numbered-Style}
%\bibliographystyle{./chapitre/acs}
%\bibliographystyle{./chapitre/IEEEtran}
%\bibliographystyle{unsrtnat}

\bibliographystyle{plainnat}
\bibliography{deep_learning}

\nopagebreak
\begin{ack}[ACKNOWLEDGMENTS]

I would like to thank the PhD students (and the Masters' students) who directly or indirectly worked on the topic: Sarra Kouider, Amel Tuama, Jer\^ome Pasquet, Hasan Abdulrahman,  Lionel Pibre, Mehdi Yedroudj, Ahmad Zakaria, during the period (2015-2018). Without all of them, this chapter would never have been possible. 

I would also like to thank my two colleagues, Fr\'ed\'eric Comby and G\'erard Subsol, who helped me supervise this nice small-world.

I thank the French working group, Caroline Fontaine, Patrick Bas, R\' emi Cogranne, with whom I have many interesting discussion and who encourage me to write this chapter. I would like to thank the Direction G\'en\'erale de l'Armement (DGA) for its support on steganalysis through the Alaska project ANR (ANR-18-ASTR-0009).

I thank the LIRMM (the lab), ICAR (my team - with all the members), the Montpellier University and the N\^imes university, HPC@LR, for all the given resources which allowed me to run such a work.

Finally, I would like to thank my wife, Nathalie, my four little smurfs, Noam, Naty, Coline, Mila, and Louis, who are the guardian of my sanity $:-)$
%\source{}
\end{ack}

%%%%%%%%%%%%%%%%%%%%%%%%%%%%%%%%%%%
%% Biography
%%%%%%%%%%%%%%%%%%%%%%%%%%%%%%%%%%%
\section*{Author Biography}

\begin{minipage}{0.3\linewidth}
%\hspace{-0.2cm}
\hspace{+0.4cm}\includegraphics[width=3cm]{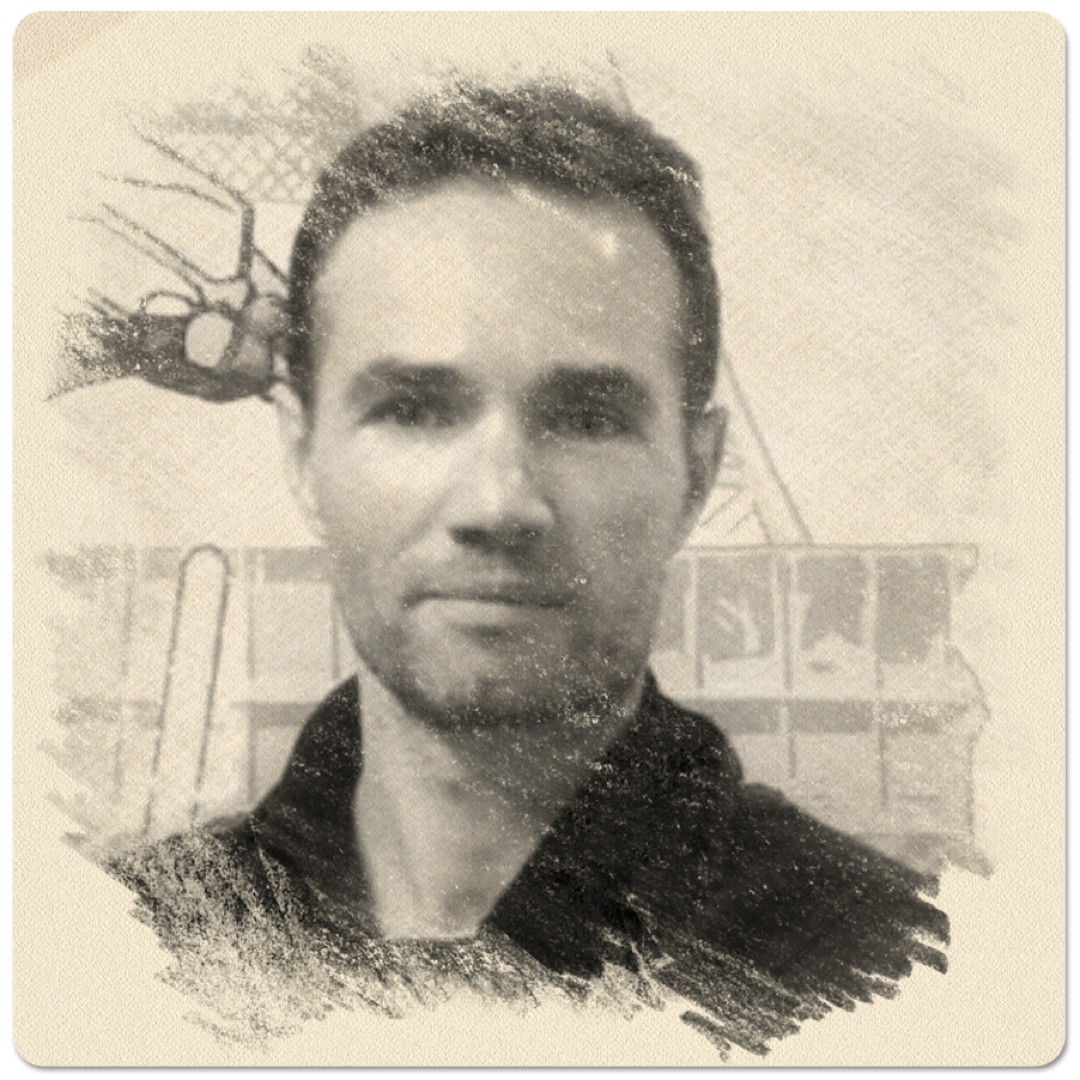}
\end{minipage}%\hfill
%\vspace{0.1cm}
\begin{minipage}{0.74\linewidth}
Marc CHAUMONT is Associate Professor (HDR Hors-Classe) accredited to supervise research, at the LIRMM laboratory (Laboratory of Computer Science, Robotics and Microelectronic), University of Montpellier and University of N\^imes, in France. He is a member of the IEEE Signal Processing - Information Forensics Security - Technical Committee and is a reviewer for the major conferen- 
\end{minipage}

\begin{minipage}{\linewidth}
ces and journals related to steganography/steganalysis. He joined the LIRMM in September 2005. He received his PhD in Computer Sciences at IRISA Rennes in 2003. His main research interests are in steganography, steganalysis, digital image forensic, and objects detection with Deep Learning.
\end{minipage}

\end{backmatter}

\end{document}